\documentclass[11pt]{article}   	% use "amsart" instead of "article" for AMSLaTeX format
\usepackage{geometry}                		% See geometry.pdf to learn the layout options. There are lots.
\pdfoutput=1

\UseRawInputEncoding

\usepackage{setspace}
\usepackage{graphicx}
\usepackage{amssymb}
\usepackage{bm}
\usepackage{epstopdf}
\usepackage{dblaccnt}
\usepackage[square,authoryear]{natbib}
\usepackage[titletoc,title]{appendix}
\usepackage[font=small,labelfont=bf]{caption}
\usepackage{placeins}

\title{Mercury's crustal thickness correlates\\ with lateral variations in mantle melt production}
\author{M. Beuthe${}^{(1)}$, B. Charlier${}^{(2)}$, O. Namur${}^{(3)}$,
A. Rivoldini${}^{(1)}$, and T. Van Hoolst${}^{(1,3)}$ \\
\small \it (1) Royal Observatory of Belgium, (2) University of Li{\`e}ge, Belgium, (3)  KU Leuven, Belgium\\
\small \it E-mail: mikael.beuthe@observatoire.be}      
\date{\small Published in Geophysical Research Letters 47 (2020) e2020GL087261}					% Activate to display a given date or no date

\begin{document}

\maketitle

%% \begin{abstract} starts the second page

\begin{abstract}
Over the first billion years of Mercury's history, mantle melting and surface volcanism produced a secondary magmatic crust varying spatially in composition and mineralogy.
By combining geochemical mapping from MESSENGER with laboratory experiments on partial melting, we translate the surface mineralogy into lateral variations of surface density and calculate the degree of mantle melting required to produce surface rocks.
If lateral density variations extend through the whole crust, the local crustal thickness correlates well with the degree of mantle melting.
Low-degree mantle melting produced a thin crust below the Northern Volcanic Plains ($19\pm3\,$km) whereas high-degree melting produced the thickest crust in the ancient High-Mg region ($50\pm12\,$km), refuting the hypothesis of an impact origin for that region.
The thickness-melting correlation has also been observed for the oceanic crust on Earth and might be a common feature of secondary crust formation on terrestrial planets.
\end{abstract}

{\center
\small\textbf{Keypoints}
\begin{enumerate}
\item Maps of surface density, degree of mantle melting, and crustal thickness from geochemical and geodetic data
\item High correlation between lateral variations in Mercury's crustal thickness and mantle melt production
\item Thick crust below ancient High-Mg region
\end{enumerate}
}

\newpage

%% ------------------------------------------------------------------------ %%
%
%  TEXT
%
%% ------------------------------------------------------------------------ %%

% section INTRODUCTION
\section{Introduction}

The surface of Mercury is mainly composed of lavas \citep{head2008}.
Major magmatic activity terminated about 3.5 Ga ago \citep{marchi2013,byrne2016} so the lavas provide a record of the early stages of the thermal and compositional evolution of the planet \citep{michel2013,tosi2013,padovan2017} and can be used to constrain melting conditions of the mantle \citep{malavergne2010,namur2016a}.
The volcanic crust of Mercury is made up of several geological units with different ages and compositions \citep{weider2015}.
Smooth plains represent the youngest units and cover about 27\% of the surface \citep{denevi2013}.
They contain the Northern Volcanic Plains (NVP), which represent the largest area of flood basalts on Mercury and possibly in the Solar System \citep{head2011}.
Recent estimates of the average thickness of Mercury's crust range from 20 to 50 km \citep{mazarico2014,padovan2015,james2015,phillips2018,sori2018,genova2019,konopliv2020}, and the mantle is ~400 km thick \citep{hauck2013,rivoldini2013,margot2018}.
The crust therefore accounts for between 6\% and 16\% of the bulk silicate content, suggesting that crustal production is as high or higher than on other terrestrial planets \citep{padovan2015,tosi2019}.
In this study we combine the most recent geochemical surface measurements, used to derive crustal density and estimates of degrees of mantle melting, with gravity and topography data obtained by the Mercury Surface, Space Environment, Geochemistry, and Ranging (MESSENGER) spacecraft.
In the northern hemisphere, where the resolution of MESSENGER geochemical and geophysical data is the highest, we identify a clear relation between geochemical units, crustal thickness, and mantle melt production.

% section METHODS
\section{Methods}

% section 2.1
\subsection{Crustal Density}

\begin{figure}
\includegraphics[width=\textwidth]{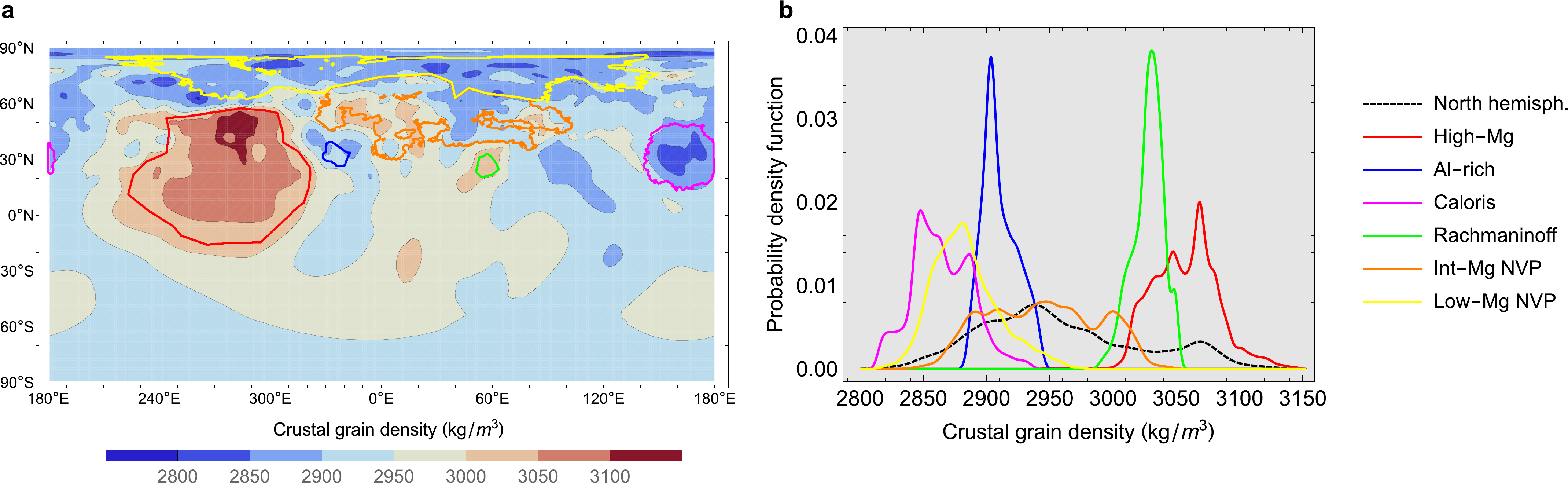}
\caption{
(a) Density of pore-free surface rocks obtained from global mineralogical mapping using MESSENGER XRS measurements for major elements (equirectangular projection).
Color lines delimitate geochemical terranes \citep{weider2015}.
(b) Histogram (equal-area) of crustal grain density for the north hemisphere and the geochemical terranes.}
\label{figdensity}
\end{figure}

Six geochemical terranes were identified on Mercury on the basis of their Al/Si and Mg/Si ratios \citep{weider2015}: the High-Mg region located in the western hemisphere at mid-northern latitudes (about 11\% of the surface), the Low-Mg NVP and Int-Mg NVP (intermediate in their Mg content) making up the Northern Volcanic Plains (about 7\% of the surface), the Rachmaninoff and Caloris basins (about 2\% of the surface), and the small Al-rich smooth plains.
These terranes make up less than half of the north hemisphere; the remaining surface is not identified with specific sets of geochemical characteristics.
Compositional variations across geochemical terranes reflect different mineralogical assemblages of surface rocks \citep{stockstillcahill2012,namur2017,vanderkaaden2017}.
Crystallization experiments on synthetic samples representative of the geochemical provinces of Mercury and mass balance calculations provide the overall distribution of surface minerals \citep{namur2017}.
Variable mineral proportions translate into pore-free crustal densities at the surface between 2785 and 3149$\rm\,kg/m^3$, with an average of 2957$\rm\,kg/m^3$ (Fig.~\ref{figdensity}; Text S1 and Fig.~S2).
This range is consistent with crustal densities previously calculated from a limited number of normative mineralogy calculations \citep{sori2018}; high-resolution gravity data might provide other constraints in the future \citep{james2019}.
The largest crustal densities (3000 to $3150\rm\,kg/m^3$) are found in the olivine-rich and plagioclase-poor High-Mg region.
The crust is the least dense (2800 to 2950$\rm\,kg/m^3$)  in the Al-rich/Mg-poor regions such as the NVP, which are plagioclase-dominated.
The Int-Mg NVP, which contain a higher proportion of olivine and pyroxene compared to Low-Mg NVP, have intermediate density values (2850 to 3050$\rm\,kg/m^3$).
If the crust is porous as on the Moon \citep{wieczorek2013} and on Mars \citep{goossens2017}, the actual crustal density is lower than what the pore-free mineralogy tells us.
We thus adopt a compaction model with porosity decreasing exponentially with increasing depth to relate the crustal density to the surface rock density \citep{besserer2014,han2014} (Text S2).
In addition, the density of rocks deeper in the crust may not vary laterally in the same way as surface rocks because magmatic flows and intrusions are not purely radial at small scale.
In particular, it is unlikely that the highest and lowest density regions extend as a straight columns from the surface down to the mantle-crust boundary.
We therefore smooth out lateral density variations with increasing depth (Text S3 and Fig.~S3).

% FIGURE 1

% section 2.2
\subsection{Crustal Thickness}

The construction of a global map of crustal thickness relies on the inversion of free-air gravity anomalies, which are the result of surface topography, relief at the mantle-crust boundary, and lateral variations in crustal and mantle densities.
For lack of high-resolution gravity data or seismic data, crustal and mantle densities are usually supposed to be laterally uniform so that gravity anomalies arise exclusively from undulations of crustal boundaries (previous work is reviewed in Text S4).
The most recent crustal thickness map for Mercury based on those assumptions has been computed from the gravity field HgM008 \citep{genova2019} and shape model GTMES 150v05 \citep{neumann2016}, which are given as spherical harmonic expansions up to degree and order 100 and 150, respectively (the spatial resolution is thus limited by the gravity field).
Here we compute a map of lateral variations in crustal thickness that takes into account lateral variations in crustal density and assess the differences with respect to a uniform density model, similarly to what has been done for the Moon \citep{wieczorek2013} and for Mars \citep{goossens2017}.
The free parameters of the model are the mantle density, surface porosity, compaction scale depth (specified by the e-folding depth), and the average crustal thickness or, alternatively, the minimum crustal thickness (Text S5).
The model also depends on the filtering degree, which parameterizes the suppression of short-wavelength gravity noise at the crust-mantle interface.
For the average crustal thickness, we adopt a default value of 35~km to facilitate comparison with previous work, but we also consider models with average thickness of 25 and 45~km, compatible with positive crust thickness everywhere.
The assumption that the surface density represents (or not) the density at depth is analyzed with the following models: a first one without smoothing, a second one with more smoothing, a third one with a laterally homogeneous lower crust below 20~km, and a fourth one for which the crustal density in the NVP is taken to be equal to the average surface density (2957$\rm\,kg/m^3$) in case the lava flows are of limited thickness compared to the total crustal thickness.
Crustal thickness could also depend on lateral variations in mantle density.
Laboratory experiments indicate that Mercury's residual mantle is mainly made up of two materials with similar densities (Fig.~S4), olivine (forsterite) and orthopyroxene (enstatite) \citep{namur2016a}.
We neglect these lateral variations and set the mantle density at the crust-mantle boundary to a uniform value of $\rho_m=3200\,\rm kg/m^3$.
This assumption could introduce errors of the order of several kilometers on the relief of the crust-mantle boundary if mantle mass anomalies are present \citep{james2015}.
We also investigate a case in which degrees 2 and 4 of the gravity and topography arise from thermal perturbations in the mantle due to uneven solar insolation \citep{tosi2015}.
This scenario requires a lithospheric thickness of more than 110~km and therefore a late capture in the 3:2 spin orbit.

% section 2.3
\subsection{Mantle Melt Production}

We calculate the degree of partial melting required to produce the observed lavas at Mercury's surface in areas for which MESSENGER XRS measured the abundance of all major elements (Ca/Si, Mg/Si, Al/Si, Na/Si, S/Si) \citep{weider2015,nittler2020}.
The variable major element chemistry -- in particular sodium -- of the lavas implies that the mantle must be slightly heterogeneous in composition, with Mg-poor lavas originating from mantle sources with lower CaO and higher $\rm Na_2O$ concentrations than Mg-rich lavas \citep{namur2016a,nittler2018}.
The two proposed mantle compositions contain similar amounts of Si, Al, and Mg. They significantly differ in Ca and Na content, but the two last components do not exceed 4 to 5 wt\% in total \citep[Table 2.2]{nittler2018}.
 As the effects of CaO and $\rm Na_2O$ on the position of the solidus (and presumably liquidus) approximately cancel out, both mantle sources have similar relations between melting fraction and temperature \citep{hirschmann2000,wasylenki2003}.
Therefore, we assume a homogeneous lherzolitic mantle (olivine and pyroxenes-bearing rock) and calculate mantle melt production using melt isopleths between the liquidus of the silicate fraction of an enstatite chondrite and the solidus of the CaO--MgO--$\rm Al_2O_3$--$\rm SiO_2$--$\rm Na_2O$ system (see further discussion in Text S7).
Based on these results, we derive an equation giving the melting degree as a function of the two ratios for which complete mapping is available, i.e.\ Mg/Si and Al/Si (Fig.~S1).

% section RESULTS
\section{Results}

\begin{figure}
\includegraphics[width=\textwidth]{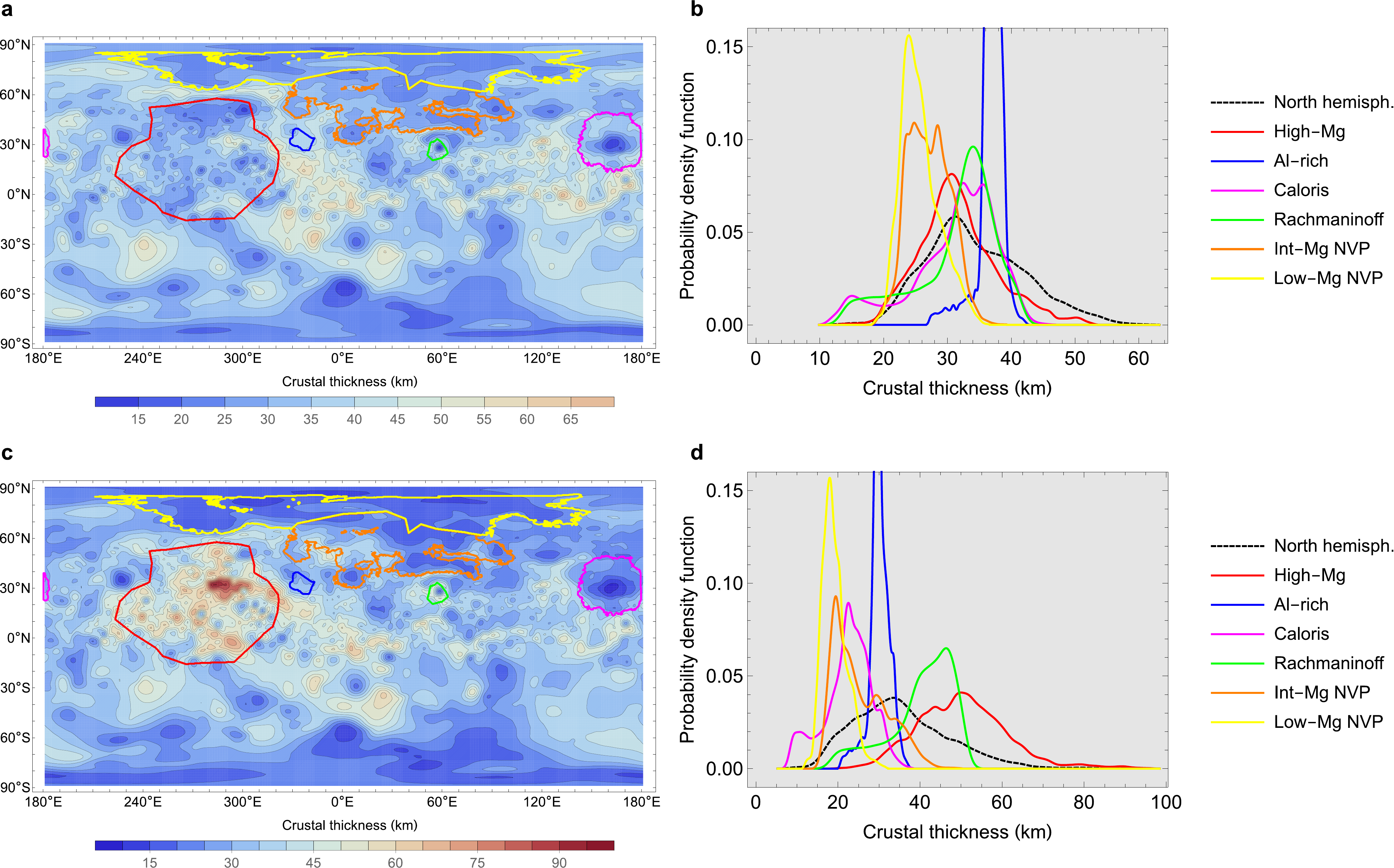}
\caption{
(a) Crustal thickness map if laterally uniform density (Model U0).
(b) Histogram of the crustal thickness in the north hemisphere for Model U0.
(c) Crustal thickness map if laterally variable density (Model V0).
(d) Histogram of the crustal thickness in the north hemisphere for Model V0.
The average crustal thickness is 35~km in both models.
Map projection is equirectangular.
Color lines delimitate geochemical terranes.
The color scale is the same in both maps.}
\label{figthickness}
\end{figure}

Crustal density variations have a significant effect on estimates of local crustal thickness (Fig.~\ref{figthickness}; Table~\ref{TableModels}; Fig. S5).
The total range of thickness variability is 60\% larger in our new reference model with non-uniform crustal density (8--97~km in our new reference Model V0; Fig.~\ref{figthickness}c,d) than without (11--67~km in Model U0 with uniform crustal density; Fig.~\ref{figthickness}a,b).
The most remarkable difference is that the High-Mg region, which has previously been interpreted as a region of thin crust, turns out to be the area with the thickest crust when crustal density variability is considered ($50\pm11\,$km; Fig.~\ref{figthickness}c,d).
The local thickening of the crust compensates the effect of the higher density on gravity.
This phenomenon is particularly strong in the highest density patch ($25-35^\circ$N, $280-300^\circ$E).
The High-Mg region is therefore not an outcrop of the mantle exhumed by a large impact \citep{weider2015}, but is rather the result of a high degree of mantle melting  \citep{frank2017,padovan2017,namur2017}.
The crust is much thinner below the NVP ($19\pm3\,$km in the Low-Mg NVP and $25\pm7\,$km in the Int-Mg NVP for model V0).
The thinnest crust is beneath the impact-related Caloris basin ($7.8\,$km for Model V0), whereas it is relatively thicker below the Rachmaninoff basin although the basin still corresponds to a local minimum of crustal thickness.
Crustal thickness calculations for different model assumptions confirm the robustness of those results (Table~\ref{TableModels} and Text S6).

%TABLE MODELS + PARTIAL RESULTS
\begin{table}[h]\centering
%\ra{1.2}
\footnotesize
\caption{
Characteristics of crustal thickness models: input parameters, crustal thickness range, and linear correlation coefficient r between crustal thickness and degree of mantle melting.}
\begin{tabular}{@{}llrrrrrrrrr@{}}
\hline
Model$^a$  & Special feature & \multicolumn{6}{c}{Input parameters$^{b}$} & \multicolumn{2}{c}{Crustal thickness} & Correl. \\
            &                          &     ${\ell_{1/2}}$ & $\phi_s$ & $d_\phi$  & $d_c$ & $d_s$  & $n_0$ & Min & Max & r \\
                        &                          &     (-)               &     (\%)    & (km)        & (km)   & (km)    &   (-)         &  (km)         & (km)           &  \\
\hline
U0        & Default if uniform density  &  40 & 24   & 8   & 35 & 35 & 20 &  $11.2$ & $67.2$    & $-0.15$  \\
V0        & Default if variable density  &  40 & 24   & 8   & 35 & 35 & 20 & $7.8$    & $97.2$      & $0.73$   \\
V1        & Lower surface porosity        &  40 & 12   & 8      & 35 & 35 & 20 & $6.4$    & $103.0$    &  $0.68$ \\
V2        & Less porous at depth          &  40 & 24   & 4      & 35 & 35 & 20 & $6.8$    & $104.5$   & $0.74$  \\
V3        & Thin crust              		 &  40 & 24   & 8    & 25 & 35 & 20 & $3.3$    & $62.4$     &  $0.57$   \\
V4        & Thick crust              		 &  40 & 24   & 8    & 45 & 45 & 20 & $12.9$  & $132.3$   &  $0.80$  \\
V5        & No degrees 2 and 4             &  40 & 24   & 8    & 35 & 35 & 20 & $6.5$    & $89.2$     &  $0.63$  \\
V6        & No smoothing                      &  40 & 24   & 8    & 35   & 0 & -- & $7.6$    & $119.9$   &  $0.73$  \\
V7        & More smoothing                  &  40 & 24   & 8    & 35 & 35 & 0 & $-0.7$   & $67.6$      & $0.66$  \\
V8        & Homogeneous lower crust  &  40 & 24   & 8    & 35 & 20 & 0 & $4.4$    & $65.9$     &  $0.39$ \\
V9        & NVP of average density      &  40 & 24   & 8    & 35 & 35 & 20 & $7.7$    & $96.8$     &  $0.71$  \\
V10 & Higher filtering degree     &  60 & 24   & 8    & 35 & 35 & 20 & $1.5$    & $119.3$    &  $0.68$ \\
\multicolumn{11}{l}{$^a$ All: mantle density $=3200\rm\,kg/m^3$; average crustal grain density $=2957\rm\,kg/m^3$; $l_{cut}=100$ (Text S5).}\\
\multicolumn{11}{l}{$^b$ $\ell_{1/2}=$ filtering degree; $\phi_s=$ surface porosity; $d_\phi=$ compaction e-folding depth; $d_c=$ average crustal}\\
\multicolumn{11}{l}{\,\,  thickness;  $d_s=$ maximum smoothing depth; $n_0=$ cutoff degree for smoothing (Text S5).}\\
\hline
\end{tabular}
\label{TableModels}
\end{table}%

% FIGURE 2

The correspondence between high Mg concentrations and thick crust suggests a correlation between the local crustal thickness and the local degree of mantle melting, i.e. the higher the degree of melting, the thicker the crust.
In agreement with previous calculations based on fewer chemical analyses but including a larger set of chemical elements, we estimate that Mercury's surface lavas were produced by 15 to 50\% partial melting of a lherzolitic mantle source \citep{namur2016a,charlier2013}.
Fig.~S1 shows the lateral variations in partial melting of the mantle required to produce those lavas.
The average melting degree is lower for the Low-Mg NVP ($25\pm3\%$) with thin crust than for the thick High-Mg terranes ($43\pm3\%$).
It ranges from 25 to 40\% for the compositionally intermediate regions, which also have intermediate crustal thicknesses in models with variable crustal density.
This further demonstrates that the High-Mg region is not an outcrop of the mantle but formed as a consequence of the highest degree of partial melting in Mercury's mantle, which led to the intrusion and eruption of Mg-rich lavas and the formation of a thick crust.

\begin{figure}
\includegraphics[width=\textwidth]{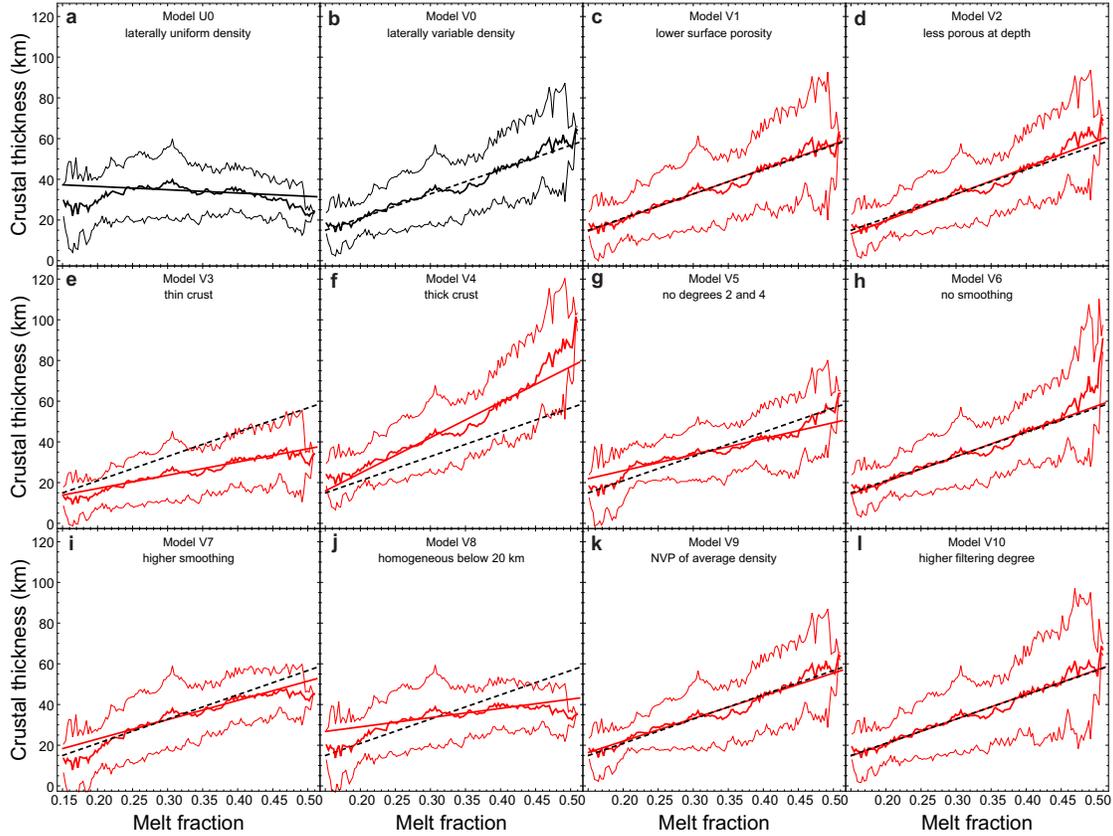}
\caption{
Crustal thickness as a function of the degree of mantle partial melting for the north hemisphere of Mercury.
Panels a-l show the 12 crustal thickness models of Table~\ref{TableModels}.
In each panel, the middle thick curve shows the moving average of the scatter plot (equal-area grid), while the top and bottom curves delimit the 95\% confidence level.
Solid straight lines represent the least-squares linear fit.
Dashed lines in panels c-l represent the linear fit for the reference Model V0 shown in panel b.
The width of melt fraction bins is equal to 0.0025.
}
\label{figcorrel}
\end{figure}

In order to quantify the statistical significance of the correlation between crustal thickness and the fraction of melt produced in the mantle, we compute the average crustal thickness and the 95\% confidence levels for melt fraction bins of equal width (Fig.~\ref{figcorrel}).
We find that crustal thickening is strongly correlated with mantle melt fraction, but only if lateral variations in crustal density are taken into account (Table~\ref{TableModels}).
In reference Model V0 (variable crustal density), the bin-averaged crustal thickness increases linearly with the bin-averaged melt fraction, with a linear correlation coefficient r = 0.73.
The coefficient of determination (r$^2$) indicates that 53\% of the crustal thickness variance can be explained by a linear dependence on the melt fraction in Model V0.
The slope and correlation between crustal thickness and mantle melt fraction depend on the values of the free parameters of the model. Lowering porosity at the surface (Model V1; r = 0.68) or within the crust (Model V2; r = 0.74) has little influence on the correlation coefficient.
Lowering the average crustal thickness to 25~km (Model V3; r = 0.57) or removing the degrees 2 and 4 (Model V5; r = 0.63) significantly decreases the correlation, whereas a higher value for the average crustal thickness increases the correlation (Model V4; r = 0.80).
Dispensing with smoothing at depth of lateral density variations (Model V6; r = 0.73) and assuming that the NVP are of average crustal density (model V9; r = 0.71) has no effect on the correlation. Increasing smoothing at depth (Model V7; r = 0.66) and increasing the filtering degree (Model V10; r = 0.68) slightly decreases the correlation although it remains significant.
The assumption that the crust is homogeneous below a depth of 20~km (Model V8; r = 0.39) results in crustal thickness variations much closer to those obtained in the model without lateral density variations so that the correlation between crustal thickness and mantle melt fraction becomes low.
A statistical analysis in the spectral domain also shows that the linear relation between crustal thickness and melt fraction is statistically significant at the longest wavelengths ($\lambda\geq\,$1400~km) for the various models considered here (Text S8; Figs. S11 and S12).

Two other issues are worth considering.
First, could correlation be biased by crust removal due to large impacts?
The answer is negative: excluding large basins \citep[Table 1]{fassett2012} has a negligible effect on the correlation unless it is already low (column r' in Table S1).
Second, correlation could be biased if lavas erupted at a late stage in the NVP because the crust was already thin in that region.
Computing the correlation without the NVP, however, barely changes the results for the models with high correlation (column r'' in Table S1).

% FIGURE 3

% section DISCUSSION
\section{Discussion}

A simple scenario of heterogeneous mantle melting followed by melt extraction and secondary crust production thus explains at least 50\% of the crustal thickness variability.
High degrees of mantle melting produced a large volume of Mg-rich lavas that accumulated to produce a thick crust.
Conversely, Mg-poorer lavas were produced by a lower degree of partial mantle melting and formed a thinner crust.
Although the degree of partial melting necessary to produce NVP lavas ($25\pm3\%$) is in the lower range for Mercury, it is relatively high as compared to most lavas produced on Earth (MORB) and is fully consistent with emplacement via flood-style volcanism. For example, the melt fraction inferred for Earth's large igneous provinces peaks at 0.25-0.30 \citep{herzberg2009}.

The production of localized thickened crustal units on Mercury might have several origins.
A main mechanism of partial melting of the mantle is the convective upwelling of hot mantle parcels \citep{michel2013,tosi2013}.
A heterogeneous pattern of upwellings and a variable degree of total partial melt might result from an early inhomogeneous production of crustal units \citep{oneill2005}, but this requirement is also in need of an explanation.
Localized high melt-fraction on Mercury might be triggered by impacts \citep{roberts2012,padovan2017}.
Alternatively, some degree of heterogeneity of the mantle composition associated with a different melt productivity, possibly with variable clinopyroxene fraction due to differentiation produced by magma ocean solidification \citep{brown2009,charlier2013,namur2016a}, will also affect the melt productivity of a mantle source with potential implications for crustal thickness \citep{katz2003}.
These explanations are partly speculative; the origin of spatial variations in partial melting remains an open question.

Crustal thickness variations unrelated to variable melt fraction can have several causes.
Although differentiation between the mantle and the secondary crust must be mainly driven by mantle melt extraction and surface emplacement, complex crustal formation processes and later modifications due to impacts and tectonics imply that the crust must be heterogeneous on small scales.
At the longest wavelengths, part of crustal thickness variability might also be an artefact of the model if lateral density variations change significantly with depth.
At shorter wavelengths, they could be due to a lack of resolution in the map of lateral density variations.

A link between crustal thickness and mantle melt production has also been identified on Earth in geodynamic settings with high production of mafic magma.
For mid-ocean ridge basalts, geochemical proxies for temperature variations in the mantle correlate with crustal thickness in the range of 2 to 15~km \citep{klein1987,asimow2001}.
The oceanic crust on Earth is also a typical example of secondary crust produced similarly to that of Mercury: adiabatic decompression and partial melting of the mantle.
There is also a large difference between the lower average thickness of the oceanic crust compared to thicker oceanic plateaus.
This is commonly explained by mantle plumes being hotter than the surrounding mantle and leading to a higher degree of mantle melting responsible for the chemical and physical characteristics of oceanic plateaus \citep{kerr2014}.
On Mars, a link has been established between the density of erupted basalts forming the crust and the temperature conditions of partial melting in the mantle \citep{baratoux2014}.
A correlation between crustal thickness, the density of the crust, and melting conditions can thus be expected.
The prime origin of these correlations is the thermal evolution of planetary interiors and particularly the cooling of the mantle over time \citep{baratoux2011,namur2016a,padovan2017}.
Variations in surface composition are thus a unique record of secular cooling of the mantle during the early history of Mercury, a planet that became magmatically inactive by 3.5 Ga \citep{byrne2016}.
Overall the correlation we observe at the planetary-scale on Mercury between mantle melt production and the thickness of the crust might be a general feature of secondary magmatic crusts of terrestrial planets.

% section CONCLUSIONS
\section{Conclusions}

Geochemical mapping of Mercury by MESSENGER, in combination with laboratory measurements, can be confidently interpreted to yield the surface distributions of rock density and of the degree of mantle melting to form these rocks.
Assuming that the composition of the crust does not change significantly with depth, we computed a new map of the crustal thickness and showed that the crust is thickest in the High-Mg terrane, previously thought to be an area of thinner crust.
This result does not support an impact origin for this terrane. In addition, more than one half of crustal thickness variations in the northern hemisphere can be explained by lateral variations in mantle melting, suggesting that heterogeneous mantle melt led to the spatially variable build-up of the crust.
The validity of the link between crustal thickness and mantle melt production will be tested in a few years by extending the analysis to the southern hemisphere when BepiColombo measurements become available.
Whatever the conclusion, the maps of surface density and partial melting of surface rocks provided here will be useful for further investigations.

\small
\section*{Acknowledgments}
We thank Mike Sori for commenting on an earlier version of the paper. We are also grateful to Nicola Tosi and Peter James for their detailed reviews.
XRS maps of Mercury are available on the Geosciences Node of the NASA Planetary Data System,
(http://pds-geosciences.wustl.edu/{\allowbreak}messenger/\allowbreak{}mess-h-xrs-3-rdr-maps-v1/{\allowbreak}messxrs\_3001/{\allowbreak}data/{\allowbreak}maps/)
and in the Supplementary Data of \citet{nittler2020}.
The gravity field HgM008 and the shape model GTMES 150v05 are also available on the Geosciences Node of the NASA Planetary Data System
(https://pds-geosciences.wustl.edu/{\allowbreak}messenger/{\allowbreak}mess-h-rss\_mla-5-sdp-v1/{\allowbreak}messrs\_1001/{\allowbreak}data/{\allowbreak}shadr/).
The maps of surface density, degree of mantle melting, and crustal thickness are available in a digital form on Zenodo (https://{\allowbreak}doi.org/{\allowbreak}10.5281/zenodo.3727115).
This research has been supported by the BRAIN-be program (BR/143/A2/COME-IN) and the Belgian PRODEX program managed by the European Space Agency in collaboration with the Belgian Federal Science Policy Office.
B.C. is a Research Associate of the Belgian Fund for Scientific Research-FNRS.
\normalsize

\newpage
%%%%%%%%%%%%%%%%%%%%%%%%%%%%%%%%%%%%%%%%%%%%%%%%%%%%%%%%%%%%%%%%%%%%%%%%%%%%
%%%%%%%%%%%%%%%%%%%%%%%%%%%%%%%%%%%%%%%%%%%%%%%%%%%%%%%%%%%%%%%%%%%%%%%%%%%%
%%%%%%%%%%%%%%%%%%%%%%%%%%%%%%%%%%%%%%%%%%%%%%%%%%%%%%%%%%%%%%%%%%%%%%%%%%%%
%%%%%%%%%%%%%%%%%%%%%%%%%%%%%%%%%%%%%%%%%%%%%%%%%%%%%%%%%%%%%%%%%%%%%%%%%%%%
%%%%%%%%%%%%%%%%%%%%%%%%%%%%%%%%%%%%%%%%%%%%%%%%%%%%%%%%%%%%%%%%%%%%%%%%%%%%

\renewcommand{\thesection}{S.\arabic{section}}
\setcounter{section}{0}
\setcounter{figure}{0}
\setcounter{table}{0}
\renewcommand{\thetable}{S.\arabic{table}}
\renewcommand{\thefigure}{S.\arabic{figure}}

%\vspace*{-2cm}
\par\noindent
%{\LARGE SUPPLEMENTARY INFORMATION \vspace{5mm}\\ Enceladus's and Dione's floating ice shells \\supported by minimum stress isostasy}
{\LARGE \bf Supporting Information for\\
``Mercury's crustal thickness correlates with lateral variations in mantle melt production"}\\

%\tableofcontents
%\listoffigures
%\listoftables
%\newpage 

%\noindent\textbf{Contents of this file}
%%%Remove or add items as needed%%%
%\begin{enumerate}
%\item Texts S1 to S8 (described in Introduction)
%\item Figures S1 to S12 (described in Introduction)
%\item Table S1. Crustal thickness models: thickness range, correlation, linear fit
%\end{enumerate}

%\clearpage

\noindent\textbf{Supporting Texts }

\begin{description} 
\item Text S1. Mineralogy and density of surface rocks
\item Text S2. Compaction model and crustal porosity
\item Text S3. Smoothing of lateral density variations with depth
\item Text S4. Lateral variations in crustal thickness: previous work
\item Text S5. Crustal thickness model: methods, benchmarking, global parameters
\item Text S6. Crustal thickness variations: assessing the role of input parameters
\item Text S7. Degree of partial melting of the mantle
\item Text S8. Correlation between melt fraction and crustal thickness
\end{description}

\vspace{5mm}

\noindent\textbf{Supporting Figures}

\begin{description} 
\item Figure S1. Degree of partial melting: map and histograms 
\item Figure S2. Errors on pore-free rock density and degree of partial melting
\item Figure S3. Depth-dependence of crustal density
\item Figure S4. Mantle density
\item Figure S5. Difference of crustal thickness between Models U0 and V0
\item Figure S6. Crustal thickness maps for Models V1 and V2
\item Figure S7. Crustal thickness maps for Models V3 and V4
\item Figure S8. Crustal thickness maps for Models V5 and V6
\item Figure S9. Crustal thickness maps for Models V7 and V8
\item Figure S10. Crustal thickness maps for Models V9 and V10
\item Figure S11. Harmonic analysis of crustal thickness and melt fraction
\item Figure S12. Harmonic analysis of correlation
\end{description}

%\begin{description}
%\item Table S1. Crustal thickness models: thickness range, correlation, linear fit
%\end{description}

\clearpage

%%%%%%%%%%%%%%%%%%%%%%%%%%%%%%%%%%%%%%%%%%%
%%%%%%%%%%%%%%%%%%%%%%%%%%%%%%%%%%%%%%%%%%%
\section{Mineralogy and density of surface rocks}

%%%%%%%%%%%%%%%%%%%%%%%%%%%%%%%%%%%%%%%%%%%%%%%%%%%%%%%%%%%%%%%%%%%%
%\noindent\textbf{Text S1. Mineralogy and density of surface rocks}

As rock samples are not available for laboratory analysis, the surface composition of the Mercurian volcanic crust is calculated using the most recent elemental composition maps produced from MESSENGER XRS data \citep{nittler2020}.
We select the pixels for which Mg/Si, Ca/Si, Al/Si and S/Si were measured.
From those data, we produce $>79,000$ compositional groups of four pixels ($0.5^\circ$ latitude $\times\,0.5^\circ$ longitude).
Each pixel group is assigned to the geochemical province in which it is located (NVP, SP, IcP-HCT, HMg), to which we attribute specific concentrations of minor elements (Ti, Mn, K) \citep{peplowski2015}.
For Na, we consider that NVP lavas have high $\rm Na_2O$ contents (Na/Si = 0.20; ca. 7~wt\% $\rm Na_2O$), SP lavas have intermediate $\rm Na_2O$ contents (Na/Si = 0.14; ca. 5~wt\% $\rm Na_2O$) and IcP-HCT and HMg lavas have lower $\rm Na_2O$ contents (Na/Si = 0.06; ca. 2~wt\% $\rm Na_2O$) \citep{peplowski2014}.
Elemental ratios are converted to oxide compositions assuming normalization to 100~wt\%.
Experiments in Fe-free systems at reducing conditions ($<\,$IW-5) show that sulphide saturation produces sulphide melt globules with an average composition of ($\rm Mg_{0.8}Ca_{0.2})S)$ \citep{namur2016b}.
We therefore recalculate S-free bulk compositions by subtracting Ca and Mg contents according to the bulk S content measured by MESSENGER before determining the silicate mineralogy of Mercurian lavas.
Additional information on the method is provided in \citet{namur2017}.

In this study, we use the silicate mineralogy of the surface of Mercury for fully solidified magma. Mineral proportions have been calculated using mass-balance between the stable liquidus phases obtained experimentally (plagioclase, quartz, forsterite, enstatite and diopside) and the bulk composition of lavas.
Detailed description of the methods is provided in \citet{namur2017}.

Calculated mineralogy is converted to pore-free crustal density using the density of minerals at $25^\circ$C.
In order to obtain a complete mapping for the density of surface rocks (including regions where Ca/Si was not measured), we fit a linear regression model to the rock density with Mg/Si and Al/Si as the only variables.
\begin{equation}
\rm 
\rho = 2.82 - 0.62(Al/Si) + 0.77(Mg/Si) + 0.40(Al/Si)^2 - 0.33(Mg/Si)^2 \hspace{5mm} (\rm{g/cm^3}) \, .
\label{eq1}
\end{equation}
The error due to the use of this equation is negligible compared to other assumptions (Fig.~\ref{figS2}; see below for a discussion of the porosity).
One sigma is $13.0\rm\,kg/m^3$ (99\% confidence is $33\rm\,kg/m^3$).

%%%%%%%%%%%%%%%%%%%%%%%%%%%%%%%%%%%%%%%%%%%%%%%%%%%%%%%%%%%%%%%%%%%%
\section{Compaction model and crustal porosity}

The crustal grain density deduced from XRS and mineralogical experiments has a (spherical) surface average of $2957\rm\,kg/m^3$.
The bulk crustal density is certainly lower than the grain density because of porosity caused by impact cratering, as has been shown for the Moon and Mars.
For the Moon, initial GRAIL gravity data implied an average porosity of 12\% to a depth of a few km, with regional variations from 4 to 21\% \citep{wieczorek2013}.
Subsequent analysis of the GRAIL extended mission data showed that the crustal density (outside mare regions) increases with wavelength, which is most easily understood if density increases with depth.
According to a simple compaction model, porosity decreases exponentially with increasing pressure \citep{han2014} or, nearly equivalently, with depth (z): $\phi(r)=\phi_s\exp(-z/d_\phi)$ where $\phi_s$ is the surface porosity; the e-folding depth $d_\phi$ is inversely proportional to the surface gravity.
For the farside highlands of the Moon, this model is best fitted by a surface porosity of 22 to 26\% and an e-folding depth of 7 to 11~km \citep{besserer2014}.
For Mars, the accumulated gravity data point to an average porosity between 10 and 23\%; it is not clear how deep this porosity extends within the crust but the negative slope of the effective density versus wavelength suggests pore closure at depth \citep{goossens2017}.

How do we extrapolate these results to Mercury?
In comparison with the Moon, the global impact flux and the mean impact velocity are larger by a factor of two \citep{lefeuvre2008}.
It does not necessarily follow that porosity is larger on Mercury.
First, the crust of the Moon's farside highlands has probably reached a fragmentation saturation threshold \citep{wiggins2019}.
Second, impacts in a highly porous crust decrease porosity \citep{milbury2015}.
Third, volcanism on Mercury may have further reduced porosity.
We thus assume that Mercury has a surface porosity of 24\% (middle value for the farside highlands of the Moon) and that it is laterally uniform (for lack of information suggesting otherwise);
we also consider a model with smaller surface porosity (12\%).
We assume that the porosity of topography is given by the surface porosity; similarly, the mantle-crust boundary (MCB) density contrast is determined by the difference between mantle and crustal densities at the average depth of the interface, where porosity is negligible.

Regarding the depth-dependence of porosity, we adopt the exponential parameterization in terms of the e-folding depth.
We do not scale the e-folding depth with the inverse of the surface gravity: such a model is applicable to compaction of sedimentary layers having the same initial porosity, whereas megaregolith results from fragmentation by impacts.
Recent numerical simulations of the development of the lunar megaregolith show that the fragmentation depth is 20\% smaller if the lunar surface gravity is replaced by its value for Mars (equal to Mercury's surface gravity), whereas it is 15\% larger if the impact velocity doubles \citep{wiggins2019}.
Combining these two factors, we estimate that the fragmentation depth is 8\% smaller on Mercury, from which we (boldly) infer a similar reduction of the compaction e-folding depth (8~km instead of 9~km).
As the latter could be further reduced by compaction, we also consider a model with a smaller e-folding depth (4~km instead of 8~km).

%%%%%%%%%%%%%%%%%%%%%%%%%%%%%%%%%%%%%%%%%%%%%%%%%%%%%%%%%%%%%%%%%%%%
\section{Smoothing of lateral density variations with depth}

The resolution of the XRS map increases from south to north with average values of $(1900, 1200, 600, 250, 200)\rm\,km$ at $(-90^\circ, 0^\circ, 30^\circ, 60^\circ, 90^\circ)$ latitude, corresponding to harmonic degrees (4; 7; 13; 30; 38) (see Fig.~4 of \citet{weider2015}).
The maps of surface mineralogy, crustal density and melt fraction (see later) have thus a coarser resolution than the gravity field, by a factor of 2 or 3. 
XRS data combined with calculation of stable mineralogy constrain the lateral variations in crustal density at the surface.
In absence of other information, we have to assume that the crustal density varies laterally in the same way deeper in the crust.
While this assumption is plausible for the large-scale density variations, it is certainly not correct at small scale.
This feature strongly increases the maximum crustal thickness for two reasons: first, the contribution of the higher density column must be compensated at the MCB and second, the MCB density contrast is very small in areas of high crustal thickness $50-100\rm\,kg/m^3$.
We thus propose to smooth lateral density variations as depth increases.
For this purpose, we use a spherical Gaussian filter of width $\beta_s=(n_0+1)^2/2$ where $n_0$ is the cutoff degree; the harmonic coefficients of the filter are computed by iteration (see \citet{jekeli2017}, pp 72-74 and pp 113-115).
The cutoff degree varies linearly from the surface ($n_0 = \ell_{max}$) to a specified degree at the MCB.
We consider models without smoothing, with moderate smoothing ($n_0=20$ at the MCB), with strong smoothing ($n_0=0$ at the MCB), and with smoothing down to a depth of 20~km (where $n_0=0$) below which the density is given by its laterally averaged value.
At the surface, the Gaussian filter has a negligible effect; at depth, it smoothens the lateral variations and restricts the density range but the mean density is unchanged (Fig.~\ref{figS3}).
By limiting the density range at depth, smoothing decreases the maximum local thickness of the crust. By including moderate smoothing in our reference model, the maximum crustal thickness is brought below 100~km.

%%%%%%%%%%%%%%%%%%%%%%%%%%%%%%%%%%%%%%%%%%%%%%%%%%%%%%%%%%%%%%%%%%%%
\section{Lateral variations in crustal thickness: previous work}

The construction of a global map of crustal thickness relies on the inversion of free-air gravity anomalies, which are due to surface topography, relief at the MCB, lateral variations in crustal and mantle densities, and CMB topography.
In the simplest model, crustal and mantle densities are laterally uniform so that gravity anomalies arise exclusively from undulations of crustal boundaries.
The average crustal thickness remains a free parameter as long as the thickness is not anchored at some location (either by seismic data or by imposing a zero minimum thickness).
This model is non-committal about the mechanical equilibrium of the crust: topography can be supported by a combination of isostasy, lithospheric flexure, and bottom loading by the mantle.
With this model, \citet{smith2012} and \citet{mazarico2014} computed the first maps of Mercury's crustal thickness based on the gravity fields HgM002 (degree and order 20) and HgM005 (degree and order 50) respectively, while \citet{genova2019} computed the latest crustal thickness map based on the gravity field HgM008 (degree and order 100).
Mercury's shape is given by the spherical harmonic solution GTMES 150v05 up to degree and order 150 \citep{neumann2016}, which has a better resolution than the gravity field.
The assumption of a laterally uniform mantle density can be relaxed by supposing that gravity anomalies and surface topography result from a combination of crustal isostasy and dynamic effects due to viscous mantle flow \citep{james2015}.
While this model showed that crustal isostasy is not the only supporting mechanism of large-scale topography (degree 13 or less), \citet{james2015} could not identify a plausible origin for long-lived deep mass anomalies in the mantle (mantle convection is of too short wavelength).
\citet{tosi2015} showed that thermal perturbations in the mantle due to uneven solar insolation could be responsible for gravity and topography at the wavelengths associated with harmonic degrees 2 and 4. 
It requires that the elastic thickness of the lithosphere was larger than 110~km at the onset of the uneven insolation, which implies that the capture into the 3:2 spin orbit resonance (causing the solar insolation pattern) occurred more than 1~Gyr after Mercury's formation.
In this work, we do not assume deep mass anomalies in the mantle but instead study the effect of lateral variations in crustal density.

%%%%%%%%%%%%%%%%%%%%%%%%%%%%%%%%%%%%%%%%%%%%%%%%%%%%%%%%%%%%%%%%%%%%
\section{Crustal thickness model: methods, benchmarking, global parameters}

\subsection*{Methods}

The computation of global crustal thickness follows a well-established procedure \citep{wieczorek1998,wieczorek2013} which is implemented with the software SHTOOLS \citep{wieczorek2018a,wieczorek2018b}.
First, free-air potential anomalies are transformed into Bouguer potential anomalies by subtracting the gravity contribution of surface topography (including finite amplitude effects).
If lateral variations in crustal density are present, their gravity contribution at the surface is removed from the Bouguer anomalies (layer by layer if density varies with depth).
Next, the remaining gravity anomaly is continued downward through the crust.
We apply a minimum amplitude low-pass filter in order to remove the short wavelengths resulting from data noise and from physical processes ignored by the model (this step can be omitted if the input fields are truncated, see below).
Finally, one obtains an equation relating the downward-continued gravity to the product of the density contrast and the MCB relief.
As this equation is nonlinear because of finite amplitude effects, it is solved for the bottom crustal relief by iteration for assumed MCB density contrasts.

The resolution of the crustal thickness map is limited by the resolution of the gravity field, which is inferior to what could be inferred from the maximum degree $\ell_{max}=100$ of its spherical harmonic expansion.
The actual resolution is estimated with the space-dependent degree strength which is defined at each point as the degree at which the signal-to-noise ratio is equal to one (e.g.\ \citet{mazarico2014}).
Because of MESSENGER's elliptical orbit, the resolution is better in the north than in the south and the degree strength accordingly increases with latitude.
Filtering data noise can be done by applying a low-pass filter when continuing the gravity signal downward through the crust, as mentioned above; this filter is characterized by the degree $\ell_{1/2}$ at which its value is equal to 0.5.
Another possibility consists in truncating the input gravity and shape at a degree $\ell_{cut}<\ell_{max}$.
Whatever the filtering procedure, there is some arbitrariness in relating either $\ell_{1/2}$ or $\ell_{cut}$ to the degree strength as the latter strongly depends on latitude.

\subsection*{Benchmarking}

We performed two different kinds of benchmarks.
For the case of uniform crustal density, we computed Mercury's crustal thickness with the model of \citet{genova2019}: gravity and topography are truncated at degree $\ell_{cut}=60$ without additional filtering, the average crustal thickness is 35~km, crustal and mantle densities are 2800 and 3200 $\rm kg/m^3$ respectively, and the shape reference radius is set to 2440~km; the resulting minimum crustal thickness is slightly negative (-1.96~km at $1^\circ$ resolution).
Our crustal thickness map differs from the one of \citet{genova2019} by 40~m at most (A. Genova, private communication).

For the case of laterally varying crustal density, we replicated the four GRAIL crustal thickness models of the Moon \citep{wieczorek2013}.
Our maps differ by a few hundred meters at most from their results, which is well within the tolerance of 1~km set for the minimum thickness and for the Apollo anchoring point.

\subsection*{Global parameters}

Crustal thickness models depend both on physical parameters (average crustal thickness, mantle density, and crustal density, which we characterize by its grain density and porosity) and technical parameters (cutoff degree and the filtering degree) to filter the noise.

Since we apply a filter for downward continuation, we can set the cutoff degree to the maximum value allowed by the gravity field ($\ell_{cut}=100$).
The filtering degree should not be larger than the degree strength of the gravity field; it can be much smaller than that if short-wavelength gravity anomalies are attributed to other causes than crustal thickness variations (e.g.\ the Moon, \citet{wieczorek2013}).
In general, a lower filtering degree reduces the crustal thickness range and allows us to consider models with a smaller average thickness or a smaller crust-mantle density contrast.
For Mercury, the main problem is that the resolution is not at all uniform: the degree strength of the field HgM008 varies between 11 (south pole) and 73 (in the north pole area), with average values of (20; 30; 40; 50; 60) at ($0^\circ$; $28^\circ$; $43^\circ$; $56^\circ$; $68^\circ$) latitude, respectively (Fig. S4 of \citet{genova2019}).
Choosing a filtering degree $\ell_{1/2}=60$ (as done by \citet{genova2019}) means that errors occur below $68^\circ$ latitude, i.e. over 93\% of the northern hemisphere.
In particular, the error can be large in the Rachmaninoff ($27.4^\circ$N) and Caloris ($30.5^\circ$N) basins where the crust is thinnest and the degree strength is about 30.
As our work is not focused on the north polar area, we adopt the more conservative choice of $\ell_{1/2}=40$, though we will consider the effect of a less efficient filter ($\ell_{1/2}=60$).

If all other parameters are fixed, the average crustal thickness $d_c$ has a lower bound which can be computed by setting the minimum crustal thickness to zero.
For example, the average crustal thickness of 35 km chosen by \citet{genova2019} yields a minimum crustal thickness close to zero, given their choices for densities and filtering.
Alternatively, the average crustal thickness was constrained to $35\pm18\,$km with geoid-to-topography ratio (GTR) method \citep{padovan2015}; using the same data but another isostatic model, \citet{sori2018} constrained it to $26\pm11\,$km.
\citet{konopliv2020} computed GTRs and localized admittances for their new gravity field (degree and order 160), yielding a 23--50~km range for the crustal thickness depending on the lower harmonic degree cutoff.
The GTR method is applicable in areas where crustal isostasy is the only mechanism supporting topography.
The isostatic assumption, however, is not supported by admittance analysis \citep{phillips2018}.
Furthermore, the High-Mg terrane makes up about half of the selected areas in the northern hemisphere where \citet{padovan2015} assume isostasy.
Among other factors, it is not clear how the higher crustal density of that terrane would affect the estimate of the average crustal thickness based on GTRs.
For this study, we adopt a default value of 35~km for the average crustal thickness to facilitate comparison with previous work, but we also consider smaller values compatible with the constraint of zero minimum thickness.

At the crust-mantle boundary, the density of an olivine/pyroxene mantle varies in the range 3180 - 3220 $\rm kg/m^3$ depending on the percentage of each mineral and on the temperature profile (Fig.~\ref{figS4}).
We set the mantle density to $\rho_m = 3200\rm\,kg/m^3$.

%%%%%%%%%%%%%%%%%%%%%%%%%%%%%%%%%%%%%%%%%%%%%%%%%%%%%%%%%%%%%%%%%%%%
\section{Crustal thickness variations: assessing the role of input parameters}

\subsection*{Uniform density}

First, we discuss the case of a crust with laterally uniform density (2957 $\rm kg/m^3$).
The reference model (U0) is specified by the parameters given in Table~1 of the main text.
Five slightly different models (U1, U2, U3, U4, U10) differ by having a lower surface porosity (U1), by being less porous at depth (U2), by having a thinner (U3) or thicker (U4) crust, or by having a higher filtering degree (U10).
Except for the crustal density, the values of the input parameters for the models (U0, U1, U2, U3, U4, U10) are the same as for the corresponding models (V0, V1, V2, V3, V4, V10); see Table~1 of the main text.
In the northern hemisphere, the crust is thinnest ($<15\,$km) in the centre of Rachmaninoff and Caloris basins and in the northeastern corner of the High-Mg region.
The crust is generally thicker ($>\,$45~km) in the equatorial regions, especially in the longitude bands $90^\circ$E to $180^\circ$E and $320^\circ$E to $30^\circ$E (with a maximum of 67~km).
The crust is thinner ($\sim$25~km) in polar areas.
These features were already noted in previous crustal thickness maps \citep{smith2012,mazarico2014,james2015,genova2019}.
A notable exception to the thicker equatorial crust occurs in the High-Mg region (centred at $270^\circ$E) where the crust is relatively thin ($\sim30\pm10\,$km) and topography is low.
On that basis, \citet{weider2015} proposed that this region could be the remnant of a giant impact basin.

The lateral variations in crustal thickness depend on global parameters (filtering degree, surface porosity, and average crustal thickness).
However, they are not very sensitive to them within the parameter ranges discussed above, except for the range in crustal thickness.
Lowering the surface porosity (Model U1) or increasing the filtering degree from $\ell_{1/2}=40$ to 60 (Model U10) extends the crustal thickness range (by 10 and 20~km respectively), but the RMS difference with Model U0 is rather small ($<2\,$km).
Less porosity at depth (Model U2) has little effect on the crustal thickness.
Changing the average crustal thickness to 25~km (Model U3) or 45~km (Model U4) only slightly changes the crustal thickness range, the main effect being the 10~km global shift.

\subsection*{Non-uniform density}

Next, we switch on lateral variations in crustal density.
The reference model (V0) and ten models with other assumptions are specified in Table~1 of the main text: V1 and V2 differ by the porosity; V3 and V4 differ by the average crustal thickness; V5 differs by the removal of degrees 2 and 4 from the crustal thickness solution (relevant if these degrees are explained by uneven solar insolation, see \citet{tosi2015}); V6, V7, and V8 differ by the smoothing of lateral density variations; V9 differs by assuming average density in the NVP; V10 differs by a higher filtering degree.
The major differences between Model V0 and Model U0 are that the High-Mg region is now the area where the crust is thickest ($50\pm11\,$km) because of its higher crustal density, and that local thickness variations are larger (from a local minimum of 8~km to the global maximum of 97~km; Fig.~2 of main text and Fig.~\ref{figS5}).
The maximum crustal thickness occurs at the western edge of the topographic plateau centred at ($30^\circ$N, $290^\circ$E) in the High-Mg region.
In comparison with Model U0, the crust is thinner below the NVP ($19\pm3\,$km in the Low-Mg NVP; $25\pm7\,$km in the Int-Mg NVP) and the Caloris basin (where the global minimum of 6.5~km occurs), whereas it is relatively thicker below Rachmaninoff basin. 

Regarding global parameters, the dependence on the porosity, average crustal thickness, and filtering degree is similar to that for Model U0: lowering the surface porosity (Model V1, Fig.~\ref{figS6}), increasing the average crustal thickness (Model V4, Fig.~\ref{figS7}), or increasing the filtering degree (Model V10, Fig.~\ref{figS10}) leads to wider ranges of crustal thickness (100 to 130~km).
Conversely, decreasing the average crustal thickness (Model V3, Fig.~\ref{figS7}) results in a smaller range (60~km).
Removing harmonic degrees 2 and 4 from the crustal thickness solution (Model V5, Fig.~\ref{figS8}) mainly decreases the thickness contrast between polar and equatorial areas.
Finally, dispensing with smoothing at depth of lateral density variations (Model V6, Fig.~\ref{figS8}) significantly increases the maximum crustal thickness (to 120~km), but these high values only occur in the small area around the global maximum at ($30^\circ$N, $280^\circ$E): crustal thickness remains smaller than 100~km nearly everywhere.
If smoothing increases more quickly with depth (Model V7, Fig.~\ref{figS9}), the High-Mg region remains a region of thicker crust ($45\pm7\,$km) but the peak crustal thickness is 30~km smaller than in Model V0.

Assuming a laterally homogeneous layer at the base of the crust (Model V8, Fig.~\ref{figS9}) results in crustal thickness variations much closer to those of the model without lateral density variations.
Finally, the main effect of assuming that the NVP are of average density (Model V9, Fig.~\ref{figS10}) is to increase the crustal thickness of the Low-Mg NVP by about 5~km.
Thus all the models with laterally variable density, except V8, have the thickest crust in the High-Mg region.

%%%%%%%%%%%%%%%%%%%%%%%%%%%%%%%%%%%%%%%%%%%%%%%%%%%%%%%%%%%%%%%%%%%%
\section{Degree of partial melting of the mantle.}

The method used to calculate the degree of partial melting of the mantle is fully described in \citet{namur2016a}.
In that study we use surface compositions obtained by the MESSENGER X-Ray Spectrometer for which all major elements were available.
The conditions of mantle melting can be constrained from the pressure-temperature position of the forsterite-enstatite-liquid multiple saturation point (MSP) and an estimate of melt production in the mantle source using melt isopleths between the solidus and liquidus surfaces.

We assume that the Mercurian mantle is dominantly lherzolitic and calculated melt isopleths between the solidus and liquidus surfaces of the silicate fraction of an EH chondrite.
This assumption is supported by the very similar compositions of the pre-melting mantle end-members (a mantle parental to the low-Mg Northern Smooth Plains and a mantle parental to the High-Mg portion of the intercrater plains and heavily cratered terrain, or IcP-HCT) obtained from mass balance calculations based on MESSENGER data and using phase equilibria results (i.e. melting fractions) \cite[Table 2.2]{nittler2018}.
These authors consider the two mantles are sufficiently close to use melting fractions obtained with a single solidus and a single liquidus by \citet{namur2016a}; their methodology is thus identical to ours.
The mantle parental to the NSP has lower (MgO,CaO) and higher alkali as compared to the mantle parental to the IcP-HCT, though these differences should be considered cautiously, as we have no direct samples of the mantle and very few constraints on the surface Na content.
The opposite effect of Na and Ca on the mantle productivity results in two mantle sources with relatively similar F (melting fraction) vs temperature relationship \citep{hirschmann2000,wasylenki2003}.
The implications of these two compositions for the exact positions of the solidus and liquidus cannot be assessed quantitatively with the information we have at the moment. 
It is thus perfectly reasonable to use a single solidus for different mantle compositions.

We use the MELTS/pMELTS algorithms to calculate the pressure-temperature path of the liquidus surface and we suppose that the solidus of the Mercurian mantle can be closely approximated by the solidus of the $\rm CaO-MgO-Al_2O_3-SiO_2-Na_2O$ system.
We selected this system because it reflects the absence of iron and the high $\rm Na_2O$ content of the Mercurian mantle.
We note that the solidus of this system at 1~GPa is similar to that experimentally determined for a bulk composition of EH chondrite \citep{berthet2009} and is thus representative for a plausible solidus of the Mercurian mantle.
We then use the pressure-temperature conditions of the forsterite-enstatite-liquid MSP calculated for surface basalts to estimate the mantle melting conditions at which these magmas were produced (see \citet{namur2016a} for details).
In S-free conditions, lavas can be produced in the mantle between $0.5$ and $2.0\,$GPa at temperatures between 1350 and $1600^\circ$C.
Most lavas from NVP, SP and IcP-HCT can be explained by isobaric melting with melt fractions ranging from 0.2 to 0.4.
The High-Mg IcP-HCT lavas would require higher melt fractions ($0.4-0.5$) and melting at relatively high temperature ($1540-1600^\circ$C).

A more realistic estimate of melting conditions includes S in the lava compositions and takes into account the role of S on pressure and temperature conditions of the MSP.
The effect of sulfur on our inferred solidus and liquidus surfaces is unknown but we assume it will be limited because S in the mantle will be present as sulphide minerals with low melting temperature which will form a sulphide-rich, silicate-poor first melt.
Further heating would then produce a S-rich silicate melt.
We use our sulfur-bearing experiments to adjust the pressure and temperature conditions of MSP calculated with pMELTS for S-free starting compositions.
We assume that the dominant effect of S on the MSP position is from the formation of CaS and MgS complexes in the silicate melt.
We assume a simple linear relationship between the CaO and MgO contents of the S-free bulk compositions and the pressure and temperature changes of the MSP when S is present.
We also consider that the sulfur on Mercury's surface was transported as dissolved $\rm S^{2-}$ in the silicate melt, which is likely given the high solubility of sulfur in silicate melt at reducing conditions.
When corrected for the effect of CaS and MgS complexing, MSP of Mercury's lavas range from $0.25$ to $1.2\,$GPa and from 1300 to $1580^\circ$C, with most basalts produced at $~0.75\,$GPa ($\sim60\,$km in depth) and $~1400^\circ$C.
Melt fractions are relatively similar to those calculated for S-free compositions and range from 0.1 to 0.5, with the highest melting fractions ($>0.45$) observed for lavas of High-Mg IcP-HCT.
The formation of High-Mg IcP-HCT lavas requires very high temperatures ($1450-1580^\circ$C). 

\begin{figure}
\includegraphics[width=0.5\textwidth]{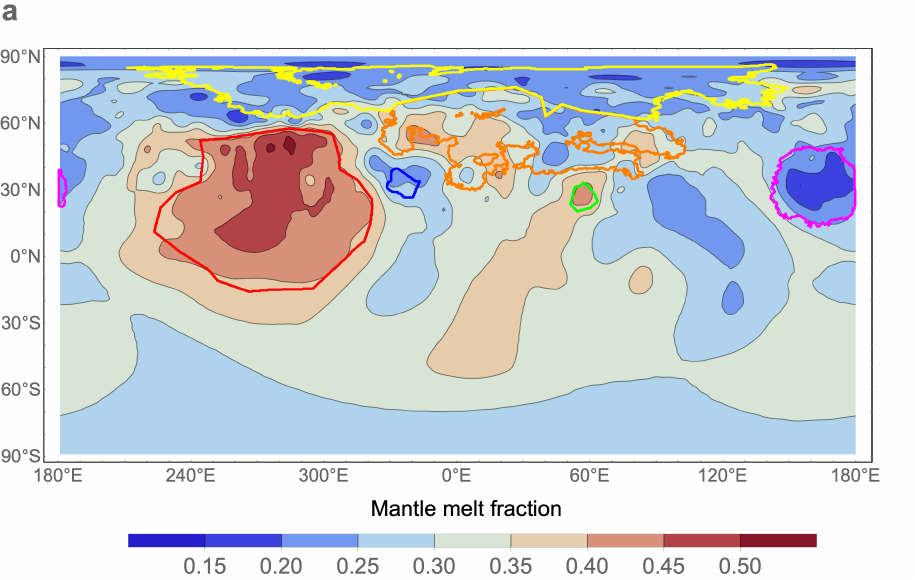}
\includegraphics[width=0.5\textwidth]{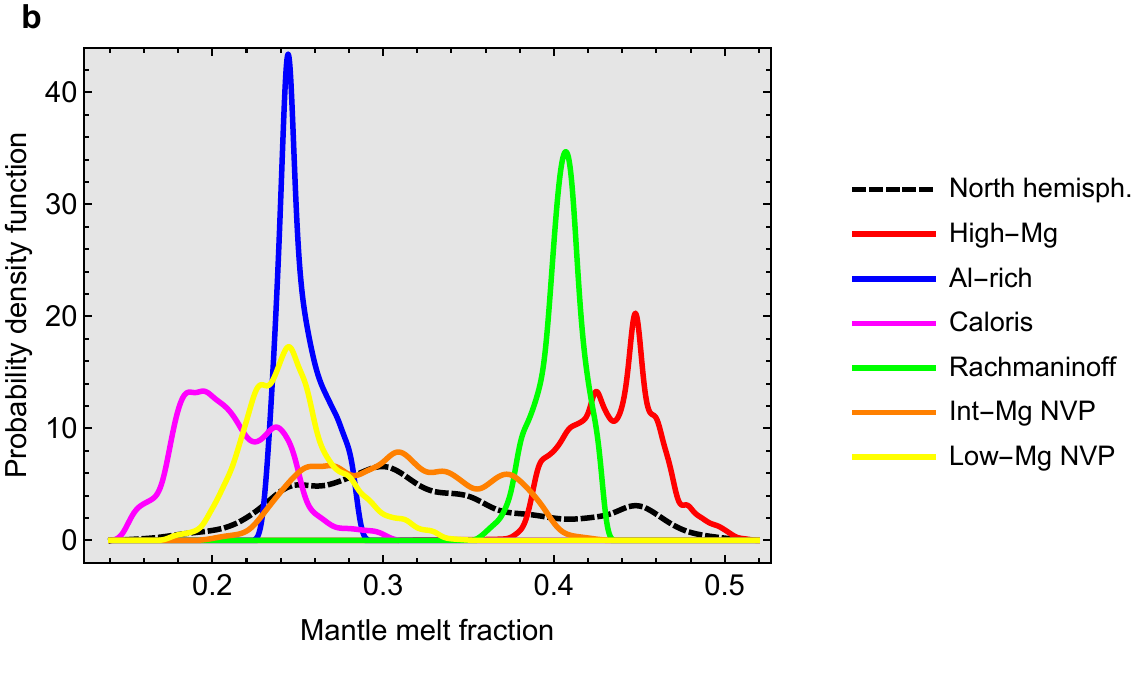}
\caption{
(a) Degree of partial melting in the mantle obtained from XRS surface compositional data for Mg/Si and Al/Si.
Color lines delimitate geochemical terranes.
(b) Histogram (equal-area) of degree of partial melting for the north hemisphere and the geochemical terranes.}
\label{figS1}
\end{figure}

\begin{figure}
 \center
 \includegraphics[width=0.7\textwidth]{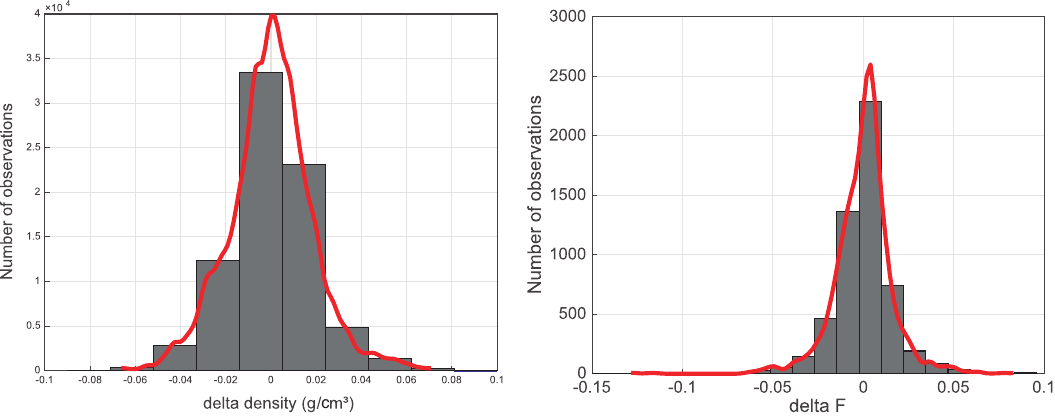}
\caption{
(left) Histogram of the difference between calculated pore-free rock densities obtained from the surface mineralogy using the full chemical dataset and densities obtained from Eq.~(\ref{eq1}) using Al/Si and Mg/Si XRS data.
(right) Histogram of the difference between the calculated degree of partial melting in the mantle using the full chemical dataset with the one obtained from Eq.~(\ref{eq2}) using Al/Si and Mg/Si XRS data in order to obtain a complete mapping of F.
Red curves represent kernel density estimates with a bandwidth equal to 0.01.
}
\label{figS2}
\end{figure}

\begin{figure}
\includegraphics[width=0.8\textwidth]{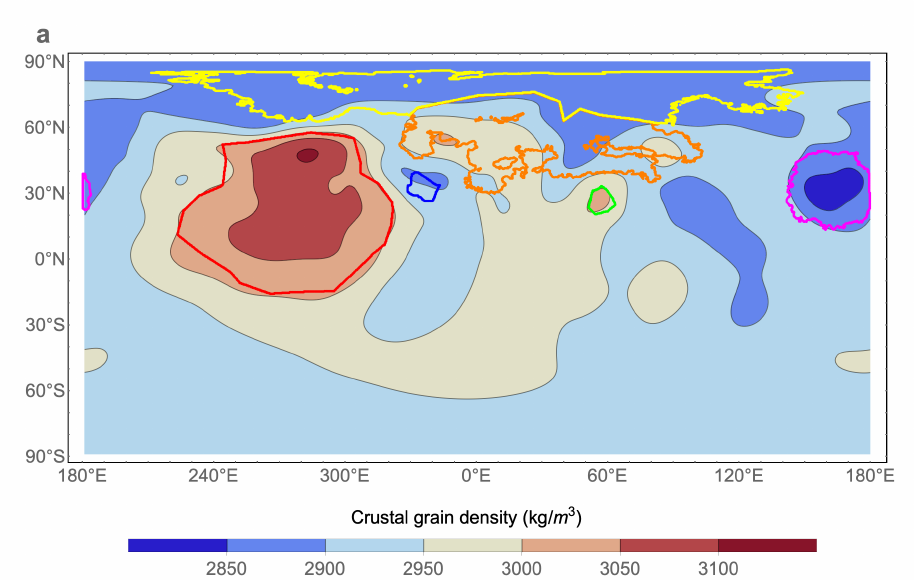}\\
\includegraphics[width=0.45\textwidth]{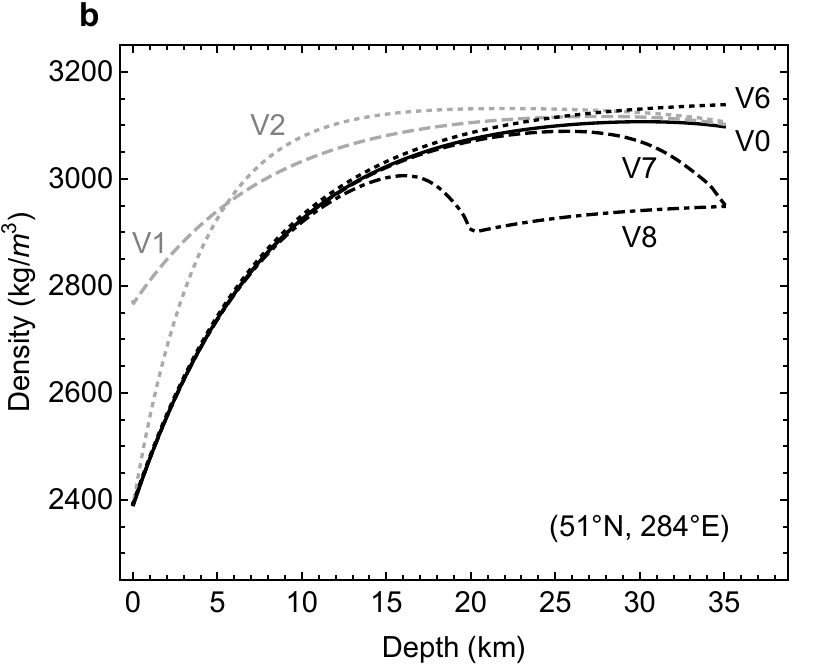}
\hspace{5mm}
\includegraphics[width=0.45\textwidth]{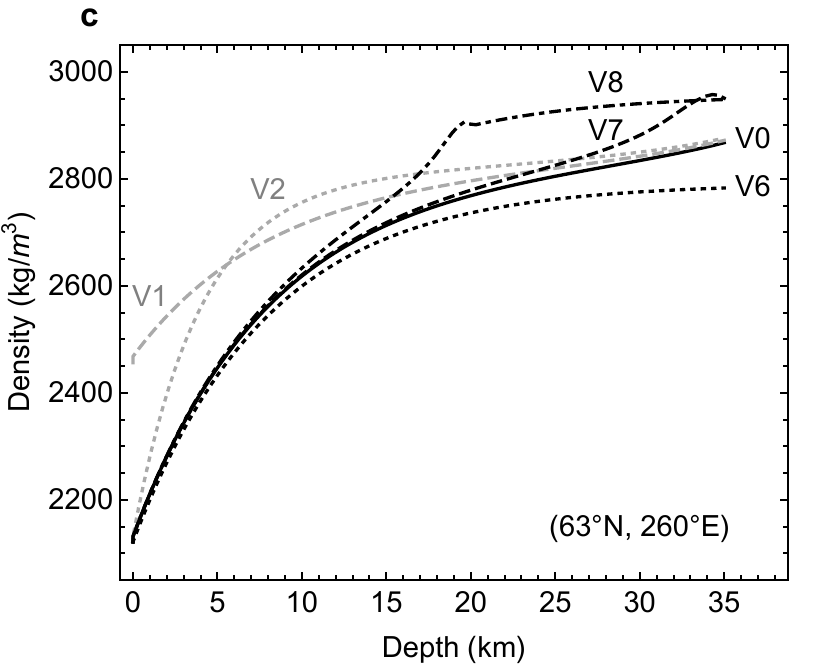}
\caption{
(a) Density of rocks at the mantle-crust boundary (depth of 35 km) for Model V0: porosity is negligible, while the effect of Gaussian smoothing is clearly visible (compare to Fig.~1).
The contours of the geochemical terranes are superimposed.
(b) Crustal density as a function of depth at the location where it is maximum at the surface.
(c) Crustal density as a function of depth at the location where it is minimum at the surface.
Density generally increases with depth because of pore closure.
Lateral smoothing (absent in V6) increases with depth and progressively pushes the density towards its average value.
If smoothing is applied up to a depth of 20 km (V8), the density below 20~km slightly increases with depth because of pore closure, but is laterally uniform.
}
\label{figS3}
\end{figure}

\begin{figure}
\includegraphics[width=0.4\textwidth]{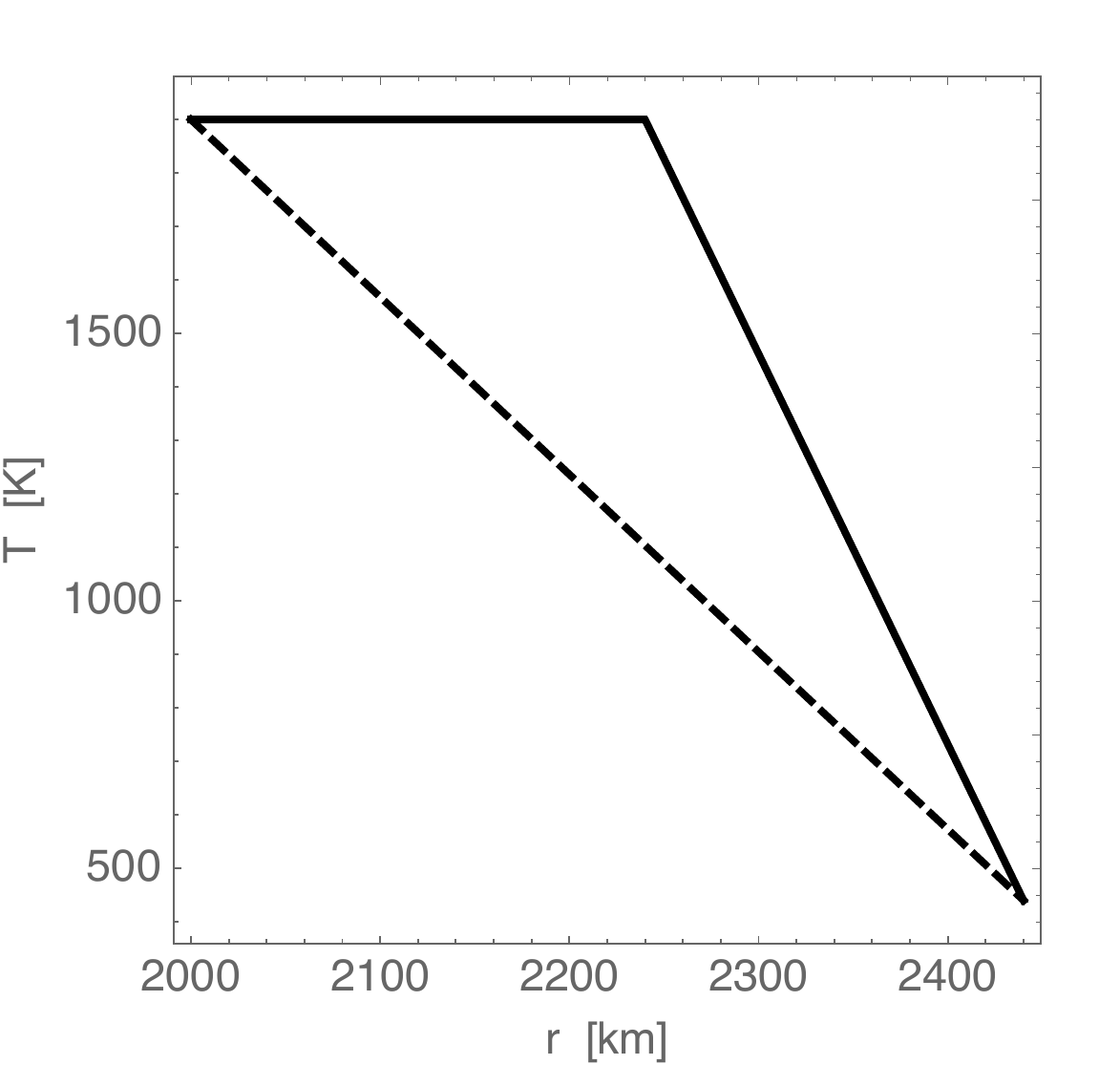}
\includegraphics[width=0.4\textwidth]{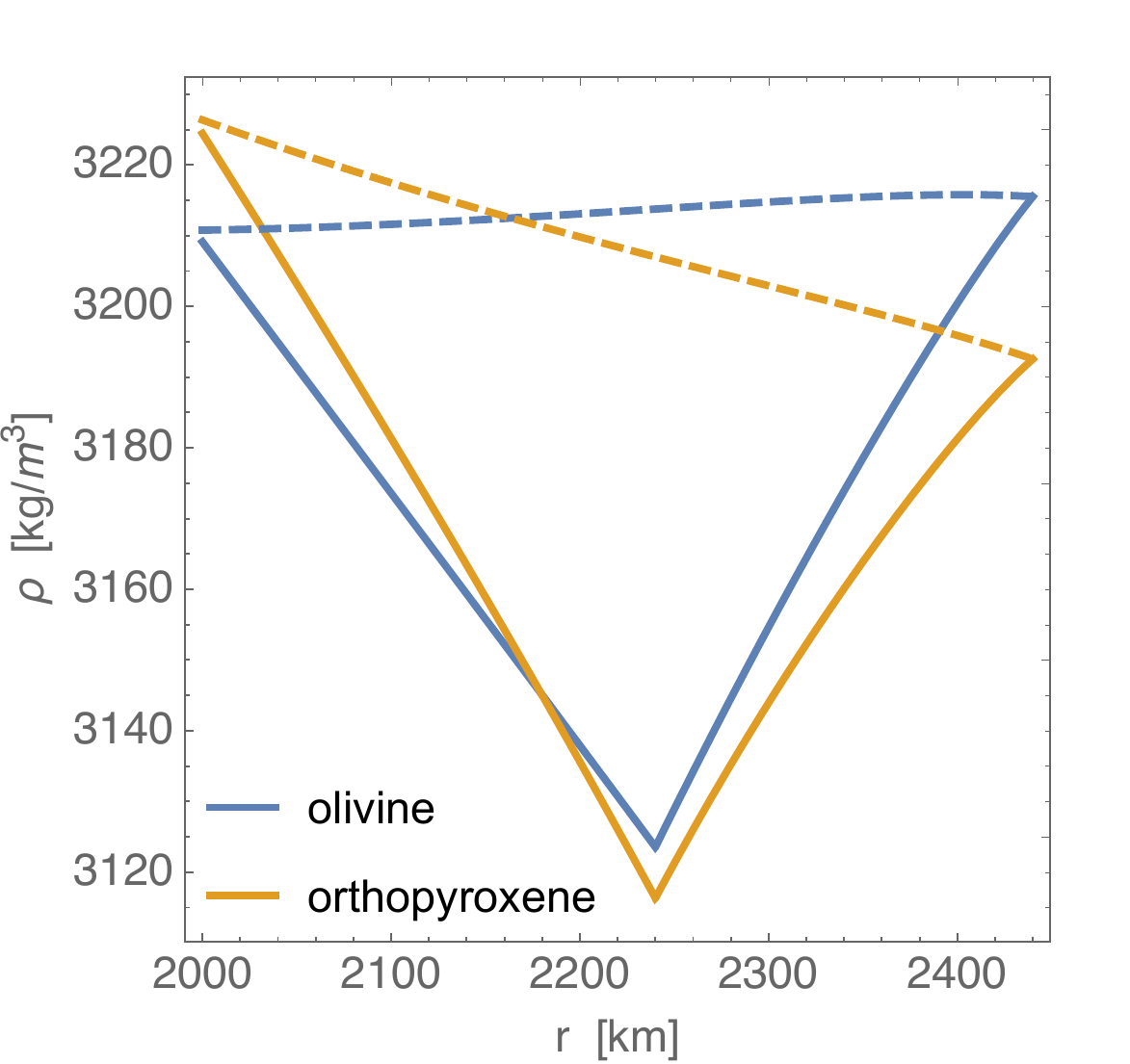}
\caption{
Mantle density: (left) temperature profiles if conductive mantle (dashed curve) or partially convective mantle (solid curve);
(right) density as a function of depth for end-member compositions of the mantle with 100\% olivine (forsterite) or 100\% orthopyroxene (enstatite).
The radius of the core is equal to 2000~km.
}
\label{figS4}
\end{figure}

\begin{figure}
\includegraphics[width=0.8\textwidth]{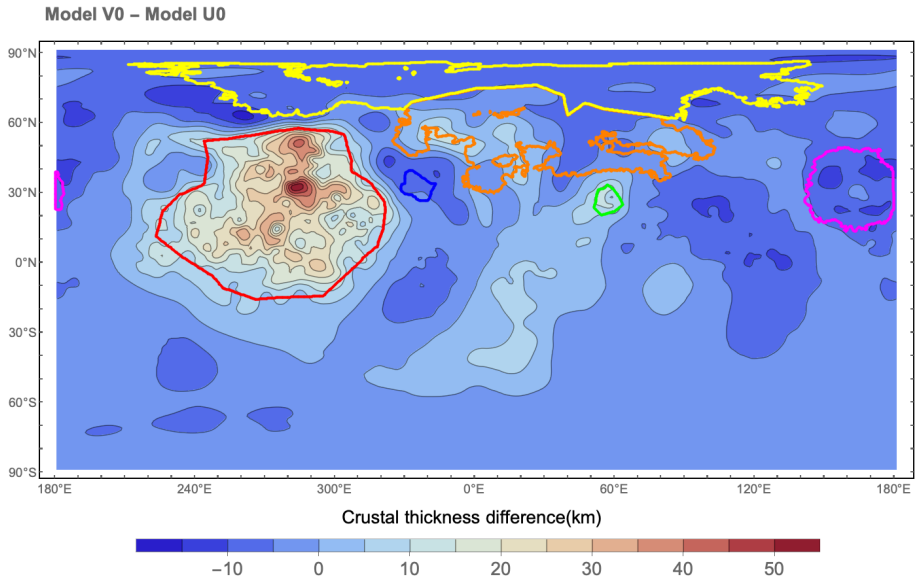}
\caption{
Map showing the difference of crustal thickness (in km) between models with laterally uniform (U0) and non-uniform (V0) crustal density.
}
\label{figS5}
\end{figure}

\begin{figure}
 \includegraphics[width=0.8\textwidth]{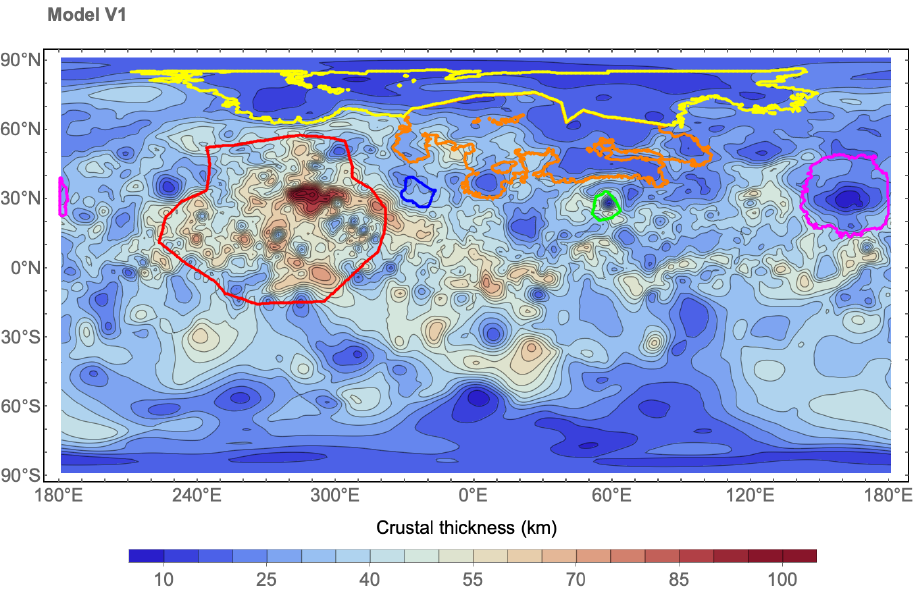}
\includegraphics[width=0.8\textwidth]{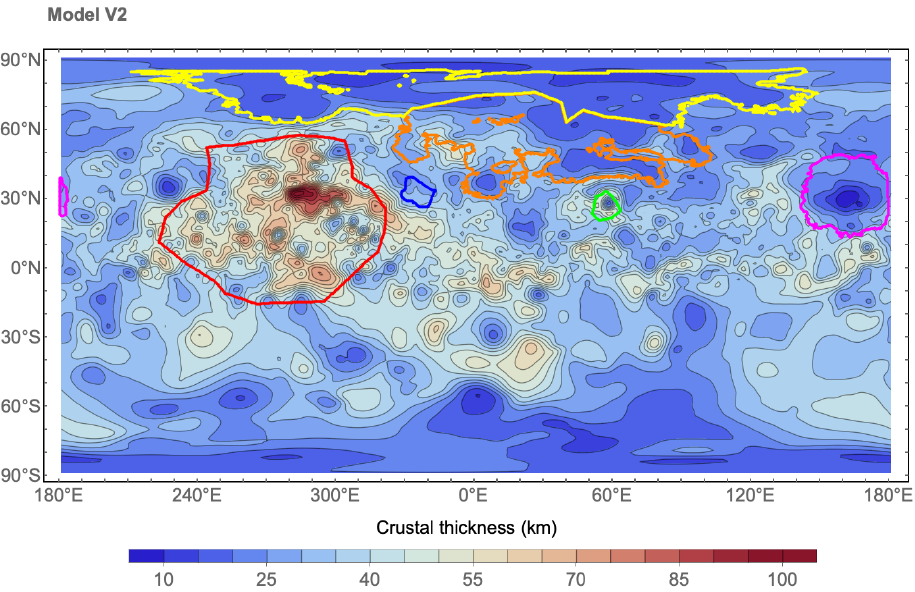}
\caption{
Crustal thickness maps for Model V1 with lower surface porosity and Model V2 with less porosity at depth.
Map projection is equirectangular.
}
\label{figS6}
\end{figure}

\begin{figure}
\includegraphics[width=0.8\textwidth]{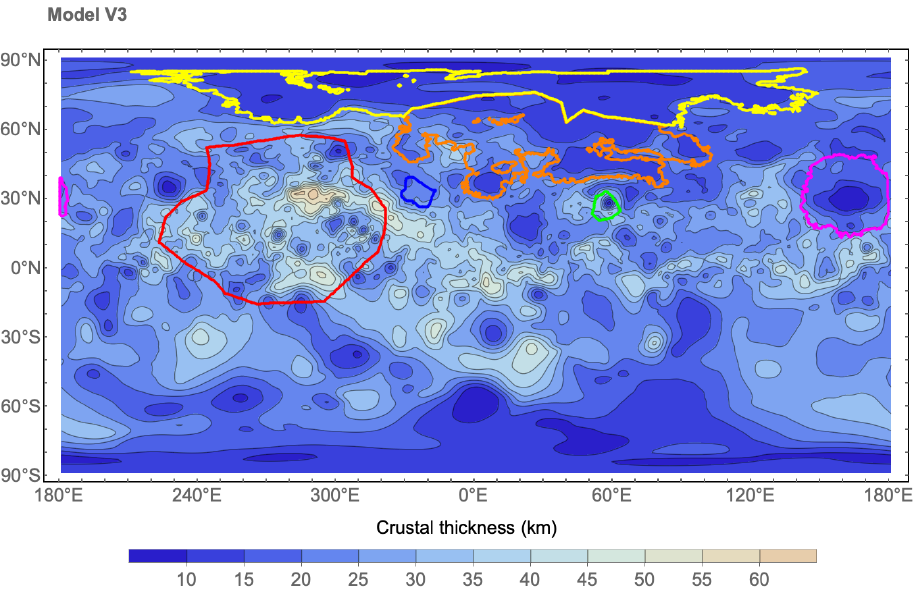}
\includegraphics[width=0.8\textwidth]{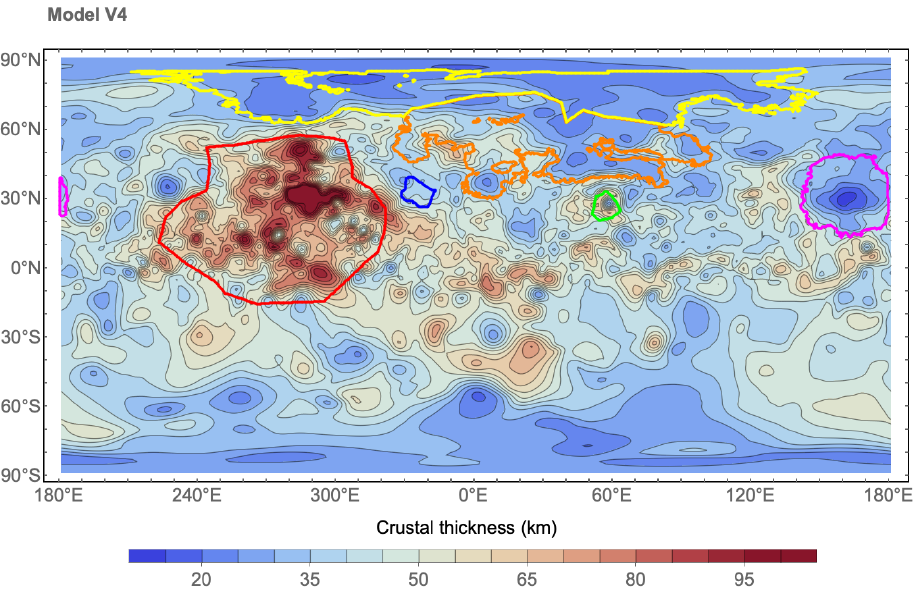}
\caption{
Crustal thickness maps for Model V3 with thinner crust and Model V4 with thicker crust.
Map projection is equirectangular.
}
\label{figS7}
\end{figure}

\begin{figure}
\includegraphics[width=0.8\textwidth]{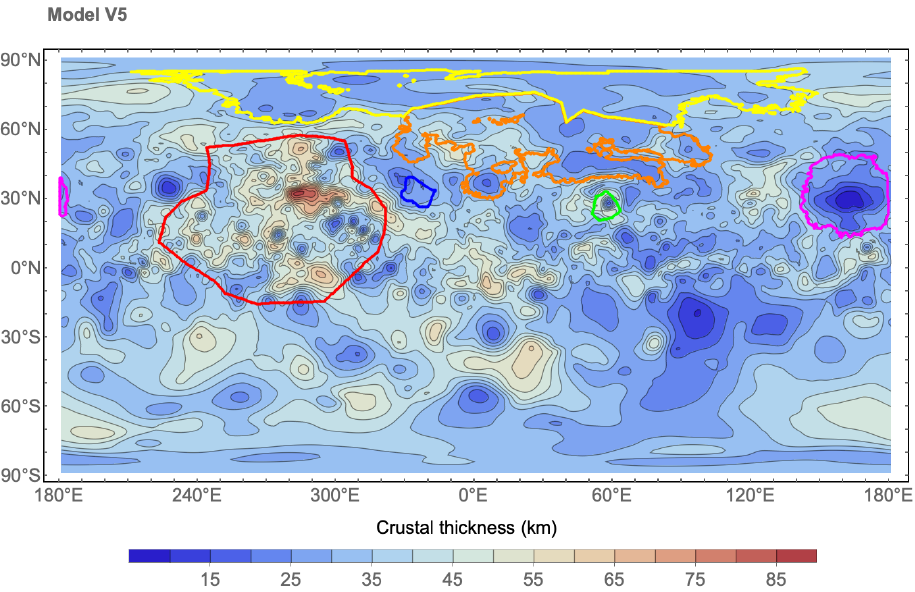}
\includegraphics[width=0.8\textwidth]{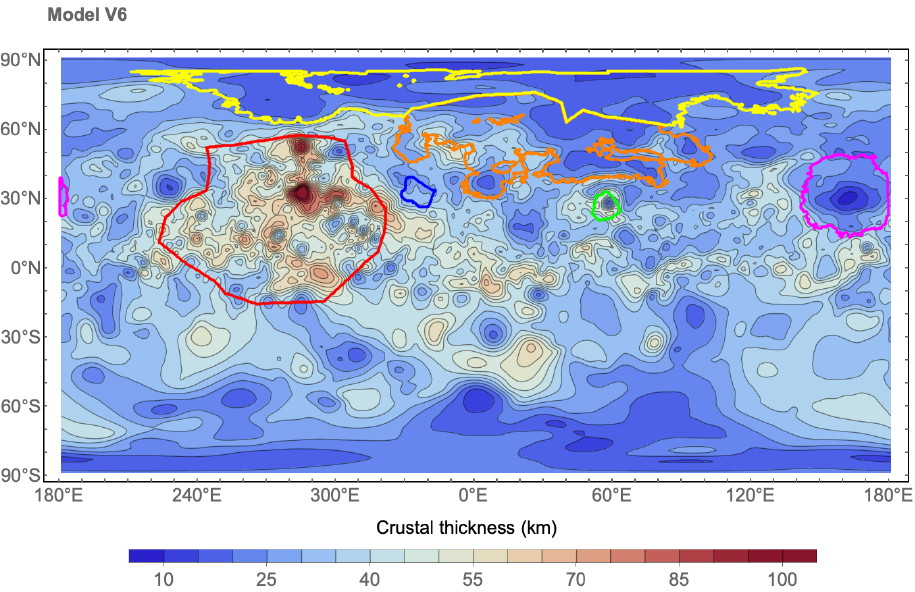}
\caption{
Crustal thickness maps for Model V5 without degrees 2 and 4 and Model V6 without smoothing.
Map projection is equirectangular.
}
\label{figS8}
\end{figure}

\begin{figure}
\includegraphics[width=0.8\textwidth]{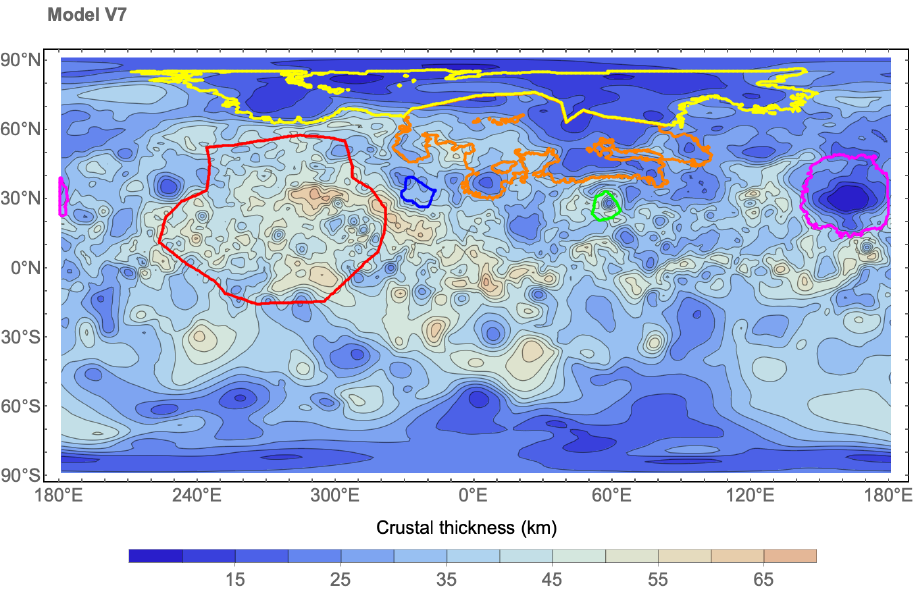}
\includegraphics[width=0.8\textwidth]{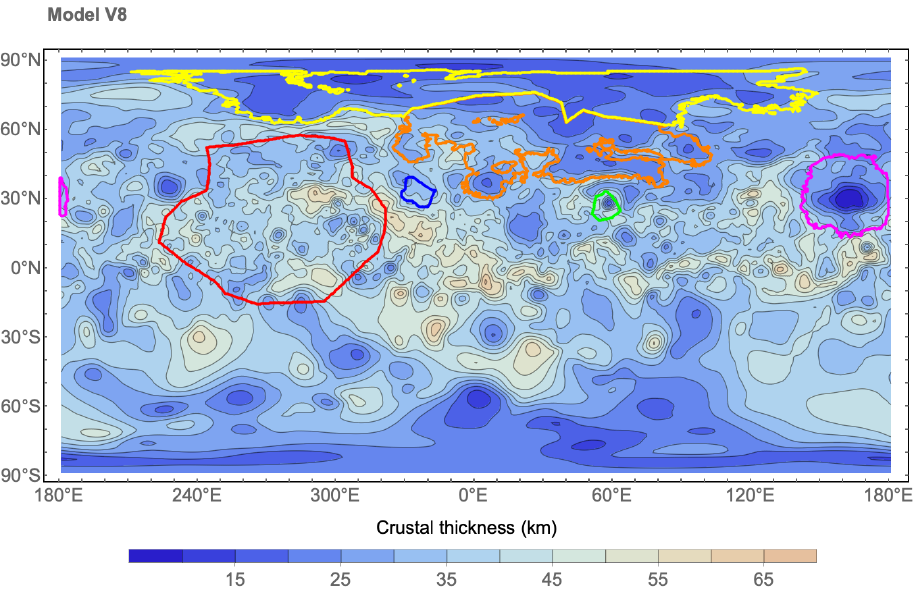}
\caption{
Crustal thickness maps for Model V7 with more smoothing and Model V8 with homogeneous lower crust.
Map projection is equirectangular.
}
\label{figS9}
\end{figure}

\begin{figure}
\includegraphics[width=0.8\textwidth]{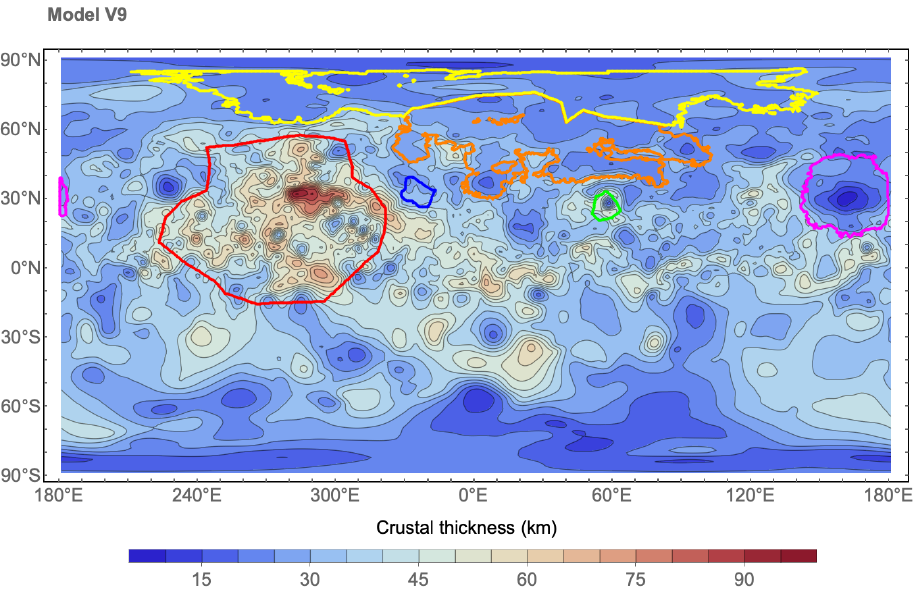}
\includegraphics[width=0.8\textwidth]{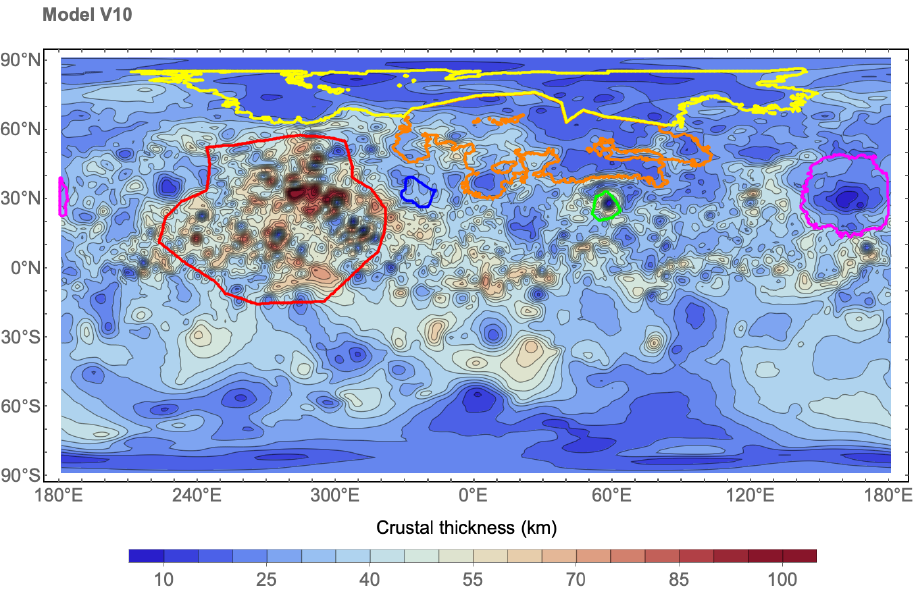}
\caption{
Crustal thickness maps for Model V9 with NVP of average density and Model V10 with higher filtering degree.
Map projection is equirectangular.
}
\label{figS10}
\end{figure}

Compositional surface maps for Mg/Si and Al/Si are complete while others (Ca/Si, S/Si) are only partial.
We thus fit a linear regression model to the degree of partial melting with Mg/Si and Al/Si as the only variables in order to obtain a complete mapping of the degree of partial melting in the mantle (Fig.~\ref{figS1}):
\begin{equation}
\rm
F = 0.08 + 0.31 (Al/Si) + 0.81 (Mg/Si) -1.75 (Al/Si)^2 - 0.34 (Mg/Si)^2 \, .
\label{eq2}
\end{equation}
The error due to the use of this equation on the degree of partial melting in the mantle is 0.011 on average and 99\% confidence is 0.039 (Fig.~\ref{figS2}).

%%%%%%%%%%%%%%%%%%%%%%%%%%%%%%%%%%%%%%%%%%%%%%%%%%%%%%%%%%%%%%%%%%%%
\section{Correlation between melt fraction and crustal thickness.}

SHTOOLS allows the user to choose between equirectangular grids and Gauss-Legendre grids.
The former makes it easier to work with MESSENGER XRS data provided as $720 \times 1440$ equirectangular grids.
Such grids, however, are not suitable for statistical analysis (regional histograms and 2D correlation analysis) because the nodes are not equally spaced on the sphere.
For that purpose, we bin the equirectangular points into a lower resolution grid with approximately equal-area cells given by an igloo pixelization (or igloo tiling) scheme \citep{crittenden1998}, in practice defined by the algorithm of \citet{leopardi2006} with either 181 or 721 latitudinal bands including polar caps.

Although the crustal thickness map must be computed globally, we interpret it only in the northern hemisphere where the resolution is highest.
In the spatial domain, it is sufficient to restrict the grid to positive latitudes.
For spectral analysis, however, one should use a smoother window in order to compute the spherical harmonic coefficients of the localized field.
The standard procedure in geophysics consists in localized spectral analysis with spherical Slepian windows \citep{wieczorek2005} coded in SHTOOLS.
We choose to use a single Slepian spherical taper with an angular radius of $90^\circ$ and a concentration factor of 0.99, corresponding to a spherical harmonic bandwidth $\Delta L = 2$.
The 1-taper window introduces a bias by reducing the amplitude of the lower latitude areas (such as the High-Mg region) in comparison with the polar regions.
The localization procedure can be slightly improved by using a larger window or multi-tapers, but we checked that this does not change the conclusion drawn below about the correlation range.

\begin{figure}
\includegraphics[width=0.5\textwidth]{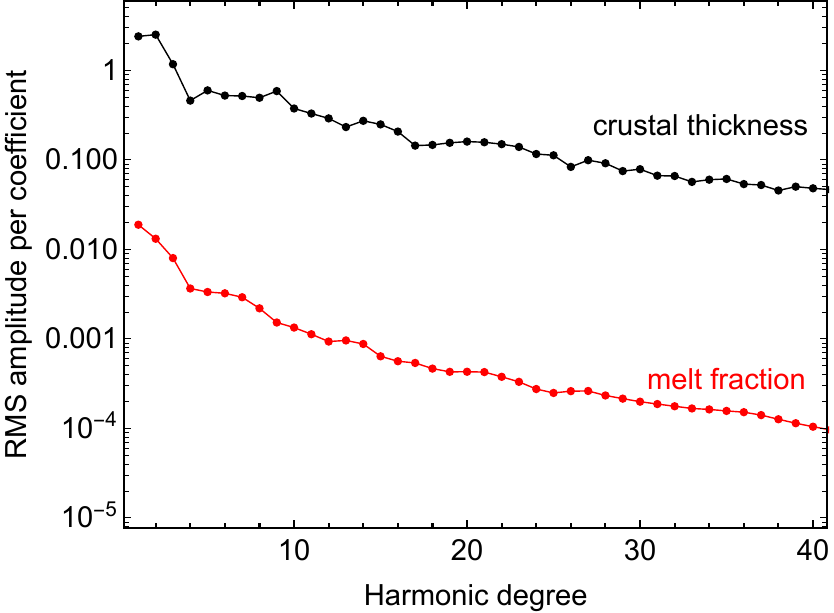}
\hspace{5mm}
\includegraphics[width=0.48\textwidth]{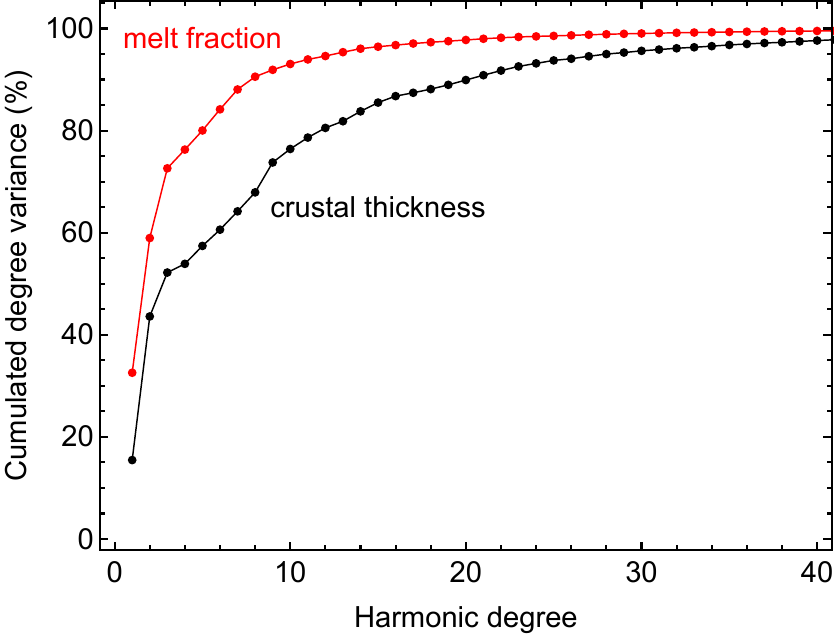}
\caption{
Harmonic analysis of crustal thickness and melt fraction (Model V0):
(left) RMS amplitude per spherical harmonic coefficient (crustal thickness is measured in km);
(right) cumulated degree variance (in \% of total variance).
}
\label{figS11}
\end{figure}

The linear correlation coefficient r cannot be submitted to statistical tests because the spectra of crustal thickness and melt fraction are dominated by long wavelengths so that the values of crustal thickness (or melt fraction) at the nodes of the equal-area grid are not independent.
Fig.~\ref{figS11} indeed shows that most power is concentrated in the first few harmonic degrees, both for the crustal thickness and the melt fraction, though the spectrum of the latter decays much faster.
The power concentration in a given harmonic range can be measured by comparing the cumulated degree variance $\sum_{\ell=1}^L\sigma_\ell$ to the total variance (given by the same formula with $L=\ell_{max}$).
For the melt fraction, the cumulated variance reaches 95\% of the total variance if $L=13$ whereas the same threshold for the crustal thickness is only reached if $L=28$ (Fig.~\ref{figS11}).
This difference is not surprising since the resolution of the melt fraction map is coarser than the gravity field by a factor of 2 or 3.

\begin{figure}
\includegraphics[width=0.5\textwidth]{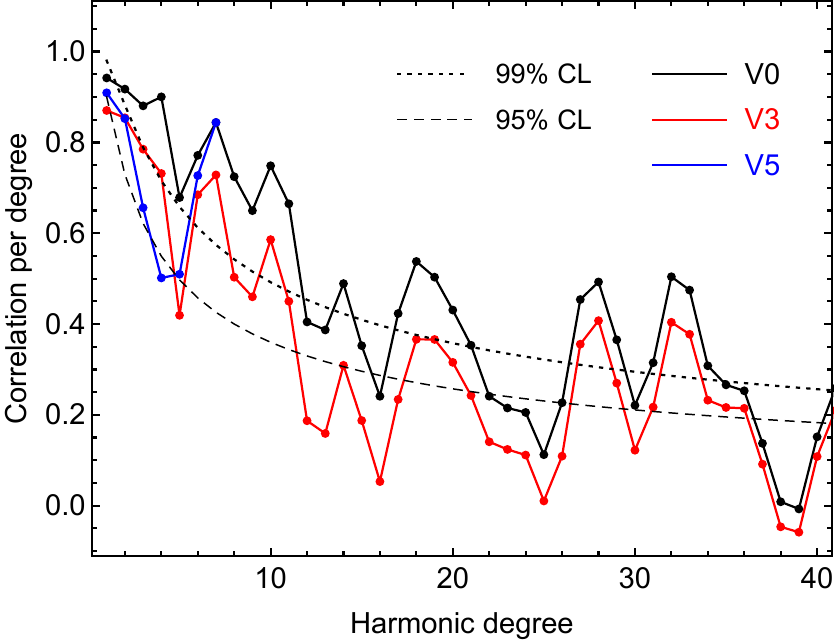}
\caption{
Harmonic analysis of correlation between melt fraction and crustal thickness (Model V0): localized spectral correlation in the northern hemisphere for Models V0, V3, and V5.
The dashed (resp. dotted) curve represents the 95\% (resp. 99\%) confidence level (CL) of positive correlation.
}
\label{figS12}
\end{figure}

The correlation of variables with skew spectra can, however, be tested in the spectral domain, where the degree correlation between random global spherical fields follows a distribution related to Student's $t$-distribution with $2\ell$ degrees of freedom \citep{eckhardt1984}.
As an additional complication, the fields must be localized in the northern hemisphere (see above) before computing the spherical harmonic coefficients; we will assume here that the statistical distribution is not biased by the localization procedure.
For Model V0, the localized spectral correlation is above the 99\% confidence level if $\ell\leq11$ (Fig.~\ref{figS12}). If the crust is thinner (Model V3), correlation drops on the whole range but is above the 95\% confidence level if $\ell\leq11$ (except marginally for $\ell=5$).
If degrees 2 and 4 are removed from the crustal thickness, correlation drops for $\ell\leq6$ (due to the coupling between degrees 2-4 and the taper with bandwidth $\Delta L=2$) but remains above the 95\% confidence level if $\ell\leq11$ (except marginally for $\ell=4$).
This analysis shows that the linear relation between crustal thickness and melt fraction is statistically significant at the longest wavelengths ($\ell\leq11$ or $\lambda\geq1400\rm\,km$) for the various models considered here, although a thin crust or the removal of degrees 2 and 4 weaken the correlation.

\clearpage

% FIGURE S1

% FIGURE S2

% FIGURE S3

% FIGURE S4

% FIGURE S5

% FIGURE S6

% FIGURE S7

% FIGURE S8

% FIGURE S9

% FIGURE S10

% FIGURE S11

% FIGURE S12

\clearpage

%TABLE RESULTS (FULL)
\begin{table}[h]\centering
%\ra{1.2}
\footnotesize
\caption{Crustal thickness range, correlation, and linear fit for all crustal thickness models.
Models Un have the same input parameters as Models Vn of Table~1, except for the density.}
\begin{tabular}{@{}lrrrrrrrrr@{}}
\hline
Model  &  \multicolumn{3}{c}{Crustal thickness$^{a}$} & \multicolumn{3}{c}{Correlation$^{b}$} & \multicolumn{3}{c}{Linear fit$^{c}$} \\
&  Min & Max & $\Delta d$ &r & r' & r'' & A & B & SD \\
            &  (km) & (km) &  (km) & (-) & (-) &(-) & (km) & (km) & (km) \\
\hline
\multicolumn{10}{l}{Uniform density}\\
U0   &  $11.2$ & $67.2$    & $0$         & $-0.15$  & $-0.26$  & $-0.29$ & n/a & n/a & n/a \\
U1   &  $7.0$   & $73.2$    & $1.6$      & $-0.15$  & $-0.25$  & $-0.29$ & n/a & n/a & n/a \\
U2   &  $10.4$ & $68.4$    &  $0.3$     & $-0.15$  & $-0.26$  & $-0.29$ & n/a & n/a & n/a \\
U3   &  $4.1$   & $52.5$    & $1.0^d$  & $-0.15$  & $-0.26$  & $-0.30$ & n/a & n/a & n/a \\
U4   &  $19.1$ & $80.9$    & $0.7^d$  & $-0.15$  & $-0.26$  & $-0.28$ & n/a & n/a & n/a \\
U10 &  $3.7$   & $79.2$    & $1.6$      & $-0.14$  & $-0.24$  & $-0.26$ & n/a & n/a & n/a \\
\multicolumn{10}{l}{Variable density}\\
V0   & $7.8$    & $97.2$    & $0$         & $0.73$   & $0.72$   & $0.72$ & $119.3$   & $-3.0$    & $8.1$  \\
V1   & $6.4$    & $103.0$  & $1.7$      &  $0.68$  & $0.66$   & $0.66$ & $122.0$   & $-3.8$    & $9.6$ \\
V2   & $6.8$    & $104.5$    & $0.8$    & $0.74$   & $0.74$   & $0.73$ & $130.3$   & $-6.4$   & $8.5$  \\
V3   & $3.3$    & $62.4$    & $3.4^d$   &  $0.57$  & $0.53$   & $0.53$ & $65.0$     & $4.1$     & $6.8$  \\
V4   & $12.9$  & $132.3$  & $3.4^d$   &  $0.80$  & $0.80$   & $0.80$ & $174.4$   & $-10.2$    & $9.4$  \\
V5   & $6.5$    & $89.2$    & $5.7$      &  $0.63$  & $0.60$   & $0.61$ & $78.8$      & $10.1$   & $7.0$  \\
V6   & $7.6$    & $119.9$  & $1.0$       &  $0.73$  & $0.73$    & $0.73$ & $123.7$   & $-4.3$    & $8.3$  \\
V7   & $-0.7$   & $67.6$    & $2.8$       & $0.66$   & $0.62$   & $0.64$ & $94.7$   & $4.1$     & $7.8$  \\
V8   & $4.4$    & $65.9$    & $5.0$       &  $0.39$  & $0.31$    & $0.31$ & $45.3$     & $20.0$   & $7.7$ \\
V9   & $7.7$    & $96.8$   & $1.1$        &  $0.71$  & $0.71$    & $0.72$ & $111.6$   & $-0.35$     & $8.0$ \\
V10 & $1.5$    & $119.3$  & $2.1$       &  $0.68$  & $0.67$    & $0.66$ & $120.5$   &  $-3.3$   & $9.5$ \\
\hline
%\multicolumn{9}{l}{${}^a$ Models Un have the same input parameters as Models Vn of Table~1, except density.}\\
\multicolumn{10}{l}{${}^a$ $\Delta d=$ RMS difference with the default model (either U0 or V0).}\\
\multicolumn{10}{l}{${}^b$ r: NH (north hemisphere);  r' = NH, no basins.; r'' = NH, no NVP.}\\
\multicolumn{10}{l}{${}^c$ A = slope; B = intercept; SD = standard deviation (whole NH).}\\
\multicolumn{10}{l}{${}^d$ If the mean difference is set to zero; otherwise $\Delta d\approx10\rm\,km$.}
\end{tabular}
\label{TableResultsFull}
\end{table}%

\bibliographystyle{agufull04}

\scriptsize
\begin{thebibliography}{65}
\providecommand{\natexlab}[1]{#1}
\expandafter\ifx\csname urlstyle\endcsname\relax
  \providecommand{\doi}[1]{doi:\discretionary{}{}{}#1}\else
  \providecommand{\doi}{doi:\discretionary{}{}{}\begingroup
  \urlstyle{rm}\Url}\fi

\bibitem[{\textit{Asimow et~al.}(2001)\textit{Asimow, Hirschmann, and
  Stolper}}]{asimow2001}
Asimow, P.~D., M.~M. Hirschmann, and E.~M. Stolper (2001), {Calculation of
  peridotite partial melting from thermodynamic models of minerals and melts,
  IV. Adiabatic decompression and the composition and mean properties of
  mid-ocean ridge basalts}, \textit{J. Petrol.}, \textit{42}(5), 963--998,
  \doi{10.1093/petrology/42.5.963}.

\bibitem[{\textit{{Baratoux} et~al.}(2011)\textit{{Baratoux}, {Toplis},
  {Monnereau}, and {Gasnault}}}]{baratoux2011}
{Baratoux}, D., M.~J. {Toplis}, M.~{Monnereau}, and O.~{Gasnault} (2011),
  {Thermal history of Mars inferred from orbital geochemistry of volcanic
  provinces}, \textit{Nature}, \textit{472}(7343), 338--341,
  \doi{10.1038/nature09903}.

\bibitem[{\textit{{Baratoux} et~al.}(2014)\textit{{Baratoux}, {Samuel},
  {Michaut}, {Toplis}, {Monnereau}, {Wieczorek}, {Garcia}, and
  {Kurita}}}]{baratoux2014}
{Baratoux}, D., H.~{Samuel}, C.~{Michaut}, M.~J. {Toplis}, M.~{Monnereau},
  M.~{Wieczorek}, R.~{Garcia}, and K.~{Kurita} (2014), {Petrological
  constraints on the density of the Martian crust}, \textit{J. Geophys. Res.},
  \textit{119}, 1707--1727, \doi{10.1002/2014JE004642}.

\bibitem[{\textit{{Berthet} et~al.}(2009)\textit{{Berthet}, {Malavergne}, and
  {Righter}}}]{berthet2009}
{Berthet}, S., V.~{Malavergne}, and K.~{Righter} (2009), {Melting of the
  Indarch meteorite (EH4 chondrite) at 1 GPa and variable oxygen fugacity:
  Implications for early planetary differentiation processes},
  \textit{Geochimica Cosmochimica Acta}, \textit{73}(20), 6402--6420,
  \doi{10.1016/j.gca.2009.07.030}.

\bibitem[{\textit{{Besserer} et~al.}(2014)\textit{{Besserer}, {Nimmo},
  {Wieczorek}, {Weber}, {Kiefer}, {McGovern}, {Andrews-Hanna}, {Smith}, and
  {Zuber}}}]{besserer2014}
{Besserer}, J., F.~{Nimmo}, M.~A. {Wieczorek}, R.~C. {Weber}, W.~S. {Kiefer},
  P.~J. {McGovern}, J.~C. {Andrews-Hanna}, D.~E. {Smith}, and M.~T. {Zuber}
  (2014), {GRAIL gravity constraints on the vertical and lateral density
  structure of the lunar crust}, \textit{J. Geophys. Res.}, \textit{41},
  5771--5777, \doi{10.1002/2014GL060240}.

\bibitem[{\textit{{Brown} and {Elkins-Tanton}}(2009)}]{brown2009}
{Brown}, S.~M., and L.~T. {Elkins-Tanton} (2009), {Compositions of Mercury's
  earliest crust from magma ocean models}, \textit{Earth Planet. Sci. Lett.},
  \textit{286}(3-4), 446--455, \doi{10.1016/j.epsl.2009.07.010}.

\bibitem[{\textit{{Byrne} et~al.}(2016)\textit{{Byrne}, {Ostrach}, {Fassett},
  {Chapman}, {Denevi}, {Evans}, {Klimczak}, {Banks}, {Head}, and
  {Solomon}}}]{byrne2016}
{Byrne}, P.~K., L.~R. {Ostrach}, C.~I. {Fassett}, C.~R. {Chapman}, B.~W.
  {Denevi}, A.~J. {Evans}, C.~{Klimczak}, M.~E. {Banks}, J.~W. {Head}, and
  S.~C. {Solomon} (2016), {Widespread effusive volcanism on Mercury likely
  ended by about 3.5 Ga}, \textit{Geophys. Res. Lett.}, \textit{43}(14),
  7408--7416, \doi{10.1002/2016GL069412}.

\bibitem[{\textit{{Charlier} et~al.}(2013)\textit{{Charlier}, {Grove}, and
  {Zuber}}}]{charlier2013}
{Charlier}, B., T.~L. {Grove}, and M.~T. {Zuber} (2013), {Phase equilibria of
  ultramafic compositions on Mercury and the origin of the compositional
  dichotomy}, \textit{Earth Planet. Sci. Lett.}, \textit{363}, 50--60,
  \doi{10.1016/j.epsl.2012.12.021}.

\bibitem[{\textit{{Crittenden} and {Turok}}(1998)}]{crittenden1998}
{Crittenden}, R.~G., and N.~G. {Turok} (1998), {Exactly azimuthal pixelizations
  of the sky}, \textit{ArXiv Astrophysics e-prints}, \textit{astro-ph/9806374}.

\bibitem[{\textit{{Denevi} et~al.}(2013)\textit{{Denevi}, {Ernst}, {Meyer},
  {Robinson}, {Murchie}, {Whitten}, {Head}, {Watters}, {Solomon}, {Ostrach},
  {Chapman}, {Byrne}, {Klimczak}, and {Peplowski}}}]{denevi2013}
{Denevi}, B.~W., C.~M. {Ernst}, H.~M. {Meyer}, M.~S. {Robinson}, S.~L.
  {Murchie}, J.~L. {Whitten}, J.~W. {Head}, T.~R. {Watters}, S.~C. {Solomon},
  L.~R. {Ostrach}, C.~R. {Chapman}, P.~K. {Byrne}, C.~{Klimczak}, and P.~N.
  {Peplowski} (2013), {The distribution and origin of smooth plains on
  Mercury}, \textit{J. Geophys. Res.}, \textit{118}(5), 891--907,
  \doi{10.1002/jgre.20075}.

\bibitem[{\textit{{Eckhardt}}(1984)}]{eckhardt1984}
{Eckhardt}, D.~H. (1984), {Correlations between global features of terrestrial
  fields}, \textit{Math. Geology}, \textit{16}, 155--171,
  \doi{10.1007/BF01032214}.

\bibitem[{\textit{{Fassett} et~al.}(2012)\textit{{Fassett}, {Head}, {Baker},
  {Zuber}, {Smith}, {Neumann}, {Solomon}, {Klimczak}, {Strom}, {Chapman},
  {Prockter}, {Phillips}, {Oberst}, and {Preusker}}}]{fassett2012}
{Fassett}, C.~I., J.~W. {Head}, D.~M.~H. {Baker}, M.~T. {Zuber}, D.~E. {Smith},
  G.~A. {Neumann}, S.~C. {Solomon}, C.~{Klimczak}, R.~G. {Strom}, C.~R.
  {Chapman}, L.~M. {Prockter}, R.~J. {Phillips}, J.~{Oberst}, and F.~{Preusker}
  (2012), {Large impact basins on Mercury: Global distribution,
  characteristics, and modification history from MESSENGER orbital data},
  \textit{J. Geophys. Res.}, \textit{117}, E00L08, \doi{10.1029/2012JE004154}.

\bibitem[{\textit{{Frank} et~al.}(2017)\textit{{Frank}, {Potter}, {Abramov},
  {James}, {Klima}, {Mojzsis}, and {Nittler}}}]{frank2017}
{Frank}, E.~A., R.~W.~K. {Potter}, O.~{Abramov}, P.~B. {James}, R.~L. {Klima},
  S.~J. {Mojzsis}, and L.~R. {Nittler} (2017), {Evaluating an impact origin for
  Mercury's high-magnesium region}, \textit{J. Geophys. Res.}, \textit{122}(3),
  614--632, \doi{10.1002/2016JE005244}.

\bibitem[{\textit{{Genova} et~al.}(2019)\textit{{Genova}, {Goossens},
  {Mazarico}, {Lemoine}, {Neumann}, {Kuang}, {Sabaka}, {Hauck}, {Smith},
  {Solomon}, and {Zuber}}}]{genova2019}
{Genova}, A., S.~{Goossens}, E.~{Mazarico}, F.~G. {Lemoine}, G.~A. {Neumann},
  W.~{Kuang}, T.~J. {Sabaka}, S.~A. {Hauck}, D.~E. {Smith}, S.~C. {Solomon},
  and M.~T. {Zuber} (2019), {Geodetic Evidence That Mercury Has A Solid Inner
  Core}, \textit{Geophys. Res. Lett.}, \textit{46}(7), 3625--3633,
  \doi{10.1029/2018GL081135}.

\bibitem[{\textit{{Goossens} et~al.}(2017)\textit{{Goossens}, {Sabaka},
  {Genova}, {Mazarico}, {Nicholas}, and {Neumann}}}]{goossens2017}
{Goossens}, S., T.~J. {Sabaka}, A.~{Genova}, E.~{Mazarico}, J.~B. {Nicholas},
  and G.~A. {Neumann} (2017), {Evidence for a low bulk crustal density for Mars
  from gravity and topography}, \textit{Geophys. Res. Lett.}, \textit{44},
  7686--7694, \doi{10.1002/2017GL074172}.

\bibitem[{\textit{{Han} et~al.}(2014)\textit{{Han}, {Schmerr}, {Neumann}, and
  {Holmes}}}]{han2014}
{Han}, S.-C., N.~{Schmerr}, G.~{Neumann}, and S.~{Holmes} (2014), {Global
  characteristics of porosity and density stratification within the lunar crust
  from GRAIL gravity and Lunar Orbiter Laser Altimeter topography data},
  \textit{Geophys. Res. Lett.}, \textit{41}, 1882--1889,
  \doi{10.1002/2014GL059378}.

\bibitem[{\textit{{Hauck} et~al.}(2013)\textit{{Hauck}, {Margot}, {Solomon},
  {Phillips}, {Johnson}, {Lemoine}, {Mazarico}, {McCoy}, {Padovan}, {Peale},
  {Perry}, {Smith}, and {Zuber}}}]{hauck2013}
{Hauck}, S.~A., J.-L. {Margot}, S.~C. {Solomon}, R.~J. {Phillips}, C.~L.
  {Johnson}, F.~G. {Lemoine}, E.~{Mazarico}, T.~J. {McCoy}, S.~{Padovan}, S.~J.
  {Peale}, M.~E. {Perry}, D.~E. {Smith}, and M.~T. {Zuber} (2013), {The curious
  case of Mercury's internal structure}, \textit{J. Geophys. Res.},
  \textit{118}(6), 1204--1220, \doi{10.1002/jgre.20091}.

\bibitem[{\textit{{Head} et~al.}(2008)\textit{{Head}, {Murchie}, {Prockter},
  {Robinson}, {Solomon}, {Strom}, {Chapman}, {Watters}, {McClintock},
  {Blewett}, and {Gillis-Davis}}}]{head2008}
{Head}, J.~W., S.~L. {Murchie}, L.~M. {Prockter}, M.~S. {Robinson}, S.~C.
  {Solomon}, R.~G. {Strom}, C.~R. {Chapman}, T.~R. {Watters}, W.~E.
  {McClintock}, D.~T. {Blewett}, and J.~J. {Gillis-Davis} (2008), {Volcanism on
  Mercury: Evidence from the First MESSENGER Flyby}, \textit{Science},
  \textit{321}(5885), 69, \doi{10.1126/science.1159256}.

\bibitem[{\textit{{Head} et~al.}(2011)\textit{{Head}, {Chapman}, {Strom},
  {Fassett}, {Denevi}, {Blewett}, {Ernst}, {Watters}, {Solomon}, {Murchie},
  {Prockter}, {Chabot}, {Gillis-Davis}, {Whitten}, {Goudge}, {Baker},
  {Hurwitz}, {Ostrach}, {Xiao}, {Merline}, {Kerber}, {Dickson}, {Oberst},
  {Byrne}, {Klimczak}, and {Nittler}}}]{head2011}
{Head}, J.~W., C.~R. {Chapman}, R.~G. {Strom}, C.~I. {Fassett}, B.~W. {Denevi},
  D.~T. {Blewett}, C.~M. {Ernst}, T.~R. {Watters}, S.~C. {Solomon}, S.~L.
  {Murchie}, L.~M. {Prockter}, N.~L. {Chabot}, J.~J. {Gillis-Davis}, J.~L.
  {Whitten}, T.~A. {Goudge}, D.~M.~H. {Baker}, D.~M. {Hurwitz}, L.~R.
  {Ostrach}, Z.~{Xiao}, W.~J. {Merline}, L.~{Kerber}, J.~L. {Dickson},
  J.~{Oberst}, P.~K. {Byrne}, C.~{Klimczak}, and L.~R. {Nittler} (2011), {Flood
  Volcanism in the Northern High Latitudes of Mercury Revealed by MESSENGER},
  \textit{Science}, \textit{333}(6051), 1853, \doi{10.1126/science.1211997}.

\bibitem[{\textit{{Herzberg} and {Gazel}}(2009)}]{herzberg2009}
{Herzberg}, C., and E.~{Gazel} (2009), {Petrological evidence for secular
  cooling in mantle plumes}, \textit{Nature}, \textit{458}(7238), 619--622,
  \doi{10.1038/nature07857}.

\bibitem[{\textit{{Hirschmann}}(2000)}]{hirschmann2000}
{Hirschmann}, M.~M. (2000), {Mantle solidus: Experimental constraints and the
  effects of peridotite composition}, \textit{Geochem. Geophys. Geosyst.},
  \textit{1}(10), 1042--26, \doi{10.1029/2000GC000070}.

\bibitem[{\textit{{James} et~al.}(2015)\textit{{James}, {Zuber}, {Phillips},
  and {Solomon}}}]{james2015}
{James}, P.~B., M.~T. {Zuber}, R.~J. {Phillips}, and S.~C. {Solomon} (2015),
  {Support of long-wavelength topography on Mercury inferred from MESSENGER
  measurements of gravity and topography}, \textit{J. Geophys. Res.},
  \textit{120}, 287--310, \doi{10.1002/2014JE004713}.

\bibitem[{\textit{{James} et~al.}(2019)\textit{{James}, {Goossens}, and
  {Mazarico}}}]{james2019}
{James}, P.~B., S.~{Goossens}, and E.~{Mazarico} (2019), {Crustal density
  estimation from line-of-sight accelerations at mercury}, in \textit{50th
  Lunar Planet. Sci. abstract 2458}.

\bibitem[{\textit{Jekeli}(2017)}]{jekeli2017}
Jekeli, C. (2017), \textit{Spectral Methods in Geodesy and Geophysics}, CRC
  Press, Boca Raton.

\bibitem[{\textit{{Katz} et~al.}(2003)\textit{{Katz}, {Spiegelman}, and
  {Langmuir}}}]{katz2003}
{Katz}, R.~F., M.~{Spiegelman}, and C.~H. {Langmuir} (2003), {A new
  parameterization of hydrous mantle melting}, \textit{Geochem. Geophys.
  Geosyst.}, \textit{4}(9), 1073, \doi{10.1029/2002GC000433}.

\bibitem[{\textit{Kerr}(2014)}]{kerr2014}
Kerr, A. (2014), Oceanic plateaus, in \textit{Treatise on Geochemistry}, edited
  by H.~D. Holland and K.~K. Turekian, 2nd ed., pp. 631--667, Elsevier, Oxford,
  \doi{10.1016/B978-0-08-095975-7.00320-X}.

\bibitem[{\textit{{Klein} and {Langmuir}}(1987)}]{klein1987}
{Klein}, E.~M., and C.~H. {Langmuir} (1987), {Global correlations of ocean
  ridge basalt chemistry with axial depth and crustal thickness}, \textit{J.
  Geophys. Res.}, \textit{92}(B8), 8089--8115, \doi{10.1029/JB092iB08p08089}.

\bibitem[{\textit{{Konopliv} et~al.}(2020)\textit{{Konopliv}, {Park}, and
  {Ermakov}}}]{konopliv2020}
{Konopliv}, A.~S., R.~S. {Park}, and A.~I. {Ermakov} (2020), {The Mercury
  gravity field, orientation, love number, and ephemeris from the MESSENGER
  radiometric tracking data}, \textit{Icarus}, \textit{335}, 113386,
  \doi{10.1016/j.icarus.2019.07.020}.

\bibitem[{\textit{{Le Feuvre} and {Wieczorek}}(2008)}]{lefeuvre2008}
{Le Feuvre}, M., and M.~A. {Wieczorek} (2008), {Nonuniform cratering of the
  terrestrial planets}, \textit{Icarus}, \textit{197}(1), 291--306,
  \doi{10.1016/j.icarus.2008.04.011}.

\bibitem[{\textit{Leopardi}(2006)}]{leopardi2006}
Leopardi, P. (2006), {A partition of the unit sphere into regions of equal area
  and small diameter}, \textit{Electron. Trans. Numer. Anal.}, \textit{25},
  309--327.

\bibitem[{\textit{{Malavergne} et~al.}(2010)\textit{{Malavergne}, {Toplis},
  {Berthet}, and {Jones}}}]{malavergne2010}
{Malavergne}, V., M.~J. {Toplis}, S.~{Berthet}, and J.~{Jones} (2010), {Highly
  reducing conditions during core formation on Mercury: Implications for
  internal structure and the origin of a magnetic field}, \textit{Icarus},
  \textit{206}(1), 199--209, \doi{10.1016/j.icarus.2009.09.001}.

\bibitem[{\textit{{Marchi} et~al.}(2013)\textit{{Marchi}, {Chapman}, {Fassett},
  {Head}, {Bottke}, and {Strom}}}]{marchi2013}
{Marchi}, S., C.~R. {Chapman}, C.~I. {Fassett}, J.~W. {Head}, W.~F. {Bottke},
  and R.~G. {Strom} (2013), {Global resurfacing of Mercury 4.0-4.1 billion
  years ago by heavy bombardment and volcanism}, \textit{Nature},
  \textit{499}(7456), 59--61, \doi{10.1038/nature12280}.

\bibitem[{\textit{{Margot} et~al.}(2018)\textit{{Margot}, {Hauck}, {Mazarico},
  {Padovan}, and {Peale}}}]{margot2018}
{Margot}, J.-L., S.~A.~I. {Hauck}, E.~{Mazarico}, S.~{Padovan}, and S.~J.
  {Peale} (2018), Mercury's internal structure, in \textit{Mercury: The View
  after MESSENGER}, edited by S.~C. {Solomon}, L.~R. {Nittler}, and B.~J.
  {Anderson}, pp. 85--113, Cambridge University Press, Cambridge,
  \doi{10.1017/9781316650684.005}.

\bibitem[{\textit{{Mazarico} et~al.}(2014)\textit{{Mazarico}, {Genova},
  {Goossens}, {Lemoine}, {Neumann}, {Zuber}, {Smith}, and
  {Solomon}}}]{mazarico2014}
{Mazarico}, E., A.~{Genova}, S.~{Goossens}, F.~G. {Lemoine}, G.~A. {Neumann},
  M.~T. {Zuber}, D.~E. {Smith}, and S.~C. {Solomon} (2014), {The gravity field,
  orientation, and ephemeris of Mercury from MESSENGER observations after three
  years in orbit}, \textit{J. Geophys. Res.}, \textit{119}, 2417--2436,
  \doi{10.1002/2014JE004675}.

\bibitem[{\textit{{Michel} et~al.}(2013)\textit{{Michel}, {Hauck}, {Solomon},
  {Phillips}, {Roberts}, and {Zuber}}}]{michel2013}
{Michel}, N.~C., S.~A. {Hauck}, S.~C. {Solomon}, R.~J. {Phillips}, J.~H.
  {Roberts}, and M.~T. {Zuber} (2013), {Thermal evolution of Mercury as
  constrained by MESSENGER observations}, \textit{J. Geophys. Res.},
  \textit{118}(5), 1033--1044, \doi{10.1002/jgre.20049}.

\bibitem[{\textit{{Milbury} et~al.}(2015)\textit{{Milbury}, {Johnson},
  {Melosh}, {Collins}, {Blair}, {Soderblom}, {Nimmo}, {Bierson}, {Phillips},
  and {Zuber}}}]{milbury2015}
{Milbury}, C., B.~C. {Johnson}, H.~J. {Melosh}, G.~S. {Collins}, D.~M. {Blair},
  J.~M. {Soderblom}, F.~{Nimmo}, C.~J. {Bierson}, R.~J. {Phillips}, and M.~T.
  {Zuber} (2015), {Preimpact porosity controls the gravity signature of lunar
  craters}, \textit{Geophys. Res. Lett.}, \textit{42}(22), 9711--9716,
  \doi{10.1002/2015GL066198}.

\bibitem[{\textit{{Namur} and {Charlier}}(2017)}]{namur2017}
{Namur}, O., and B.~{Charlier} (2017), {Silicate mineralogy at the surface of
  Mercury}, \textit{Nat. Geosci.}, \textit{10}, 9--13, \doi{10.1038/ngeo2860}.

\bibitem[{\textit{{Namur} et~al.}(2016{\natexlab{a}})\textit{{Namur},
  {Charlier}, {Holtz}, {Cartier}, and {McCammon}}}]{namur2016b}
{Namur}, O., B.~{Charlier}, F.~{Holtz}, C.~{Cartier}, and C.~{McCammon}
  (2016{\natexlab{a}}), {Sulfur solubility in reduced mafic silicate melts:
  Implications for the speciation and distribution of sulfur on Mercury},
  \textit{Earth Planet. Sci. Lett.}, \textit{448}, 102--114,
  \doi{10.1016/j.epsl.2016.05.024}.

\bibitem[{\textit{{Namur} et~al.}(2016{\natexlab{b}})\textit{{Namur},
  {Collinet}, {Charlier}, {Grove}, {Holtz}, and {McCammon}}}]{namur2016a}
{Namur}, O., M.~{Collinet}, B.~{Charlier}, T.~L. {Grove}, F.~{Holtz}, and
  C.~{McCammon} (2016{\natexlab{b}}), {Melting processes and mantle sources of
  lavas on Mercury}, \textit{Earth Planet. Sci. Lett.}, \textit{439}, 117--128,
  \doi{10.1016/j.epsl.2016.01.030}.

\bibitem[{\textit{{Neumann} et~al.}(2016)\textit{{Neumann}, {Perry},
  {Mazarico}, {Ernst}, {Zuber}, {Smith}, {Becker}, {Gaskell}, {Head},
  {Robinson}, and {Solomon}}}]{neumann2016}
{Neumann}, G.~A., M.~E. {Perry}, E.~{Mazarico}, C.~M. {Ernst}, M.~T. {Zuber},
  D.~E. {Smith}, K.~J. {Becker}, R.~E. {Gaskell}, J.~W. {Head}, M.~S.
  {Robinson}, and S.~C. {Solomon} (2016), {Mercury shape model from laser
  altimetry and planetary comparisons}, in \textit{47th Lunar Planet. Sci.
  abstract 2087}.

\bibitem[{\textit{{Nittler} et~al.}(2018)\textit{{Nittler}, {Chabot}, {Grove},
  and {Peplowski}}}]{nittler2018}
{Nittler}, L.~R., N.~L. {Chabot}, T.~L. {Grove}, and P.~N. {Peplowski} (2018),
  The chemical composition of mercury, in \textit{Mercury: The View after
  MESSENGER}, edited by S.~C. {Solomon}, L.~R. {Nittler}, and B.~J. {Anderson},
  pp. 30--51, Cambridge University Press, Cambridge,
  \doi{10.1017/9781316650684.003}.

\bibitem[{\textit{{Nittler} et~al.}(2020)\textit{{Nittler}, {Frank}, {Weider},
  {Crapster-Pregont}, {Vorburger}, {Starr}, and {Solomon}}}]{nittler2020}
{Nittler}, L.~R., E.~A. {Frank}, S.~Z. {Weider}, E.~{Crapster-Pregont},
  A.~{Vorburger}, R.~D. {Starr}, and S.~C. {Solomon} (2020), {Global
  major-element maps of Mercury from four years of MESSENGER X-Ray Spectrometer
  observations}, \textit{Icarus}, \textit{345}, 113716,
  \doi{10.1016/j.icarus.2020.113716}.

\bibitem[{\textit{{O'Neill} et~al.}(2005)\textit{{O'Neill}, {Moresi}, and
  {Lenardic}}}]{oneill2005}
{O'Neill}, C., L.~{Moresi}, and A.~{Lenardic} (2005), {Insulation and depletion
  due to thickened crust: Effects on melt production on Mars and Earth},
  \textit{Geophys. Res. Lett.}, \textit{32}(14), L14304,
  \doi{10.1029/2005GL022855}.

\bibitem[{\textit{{Padovan} et~al.}(2015)\textit{{Padovan}, {Wieczorek},
  {Margot}, {Tosi}, and {Solomon}}}]{padovan2015}
{Padovan}, S., M.~A. {Wieczorek}, J.-L. {Margot}, N.~{Tosi}, and S.~C.
  {Solomon} (2015), {Thickness of the crust of Mercury from geoid-to-topography
  ratios}, \textit{Geophys. Res. Lett.}, \textit{42}, 1029--1038,
  \doi{10.1002/2014GL062487}.

\bibitem[{\textit{{Padovan} et~al.}(2017)\textit{{Padovan}, {Tosi}, {Plesa},
  and {Ruedas}}}]{padovan2017}
{Padovan}, S., N.~{Tosi}, A.-C. {Plesa}, and T.~{Ruedas} (2017),
  {Impact-induced changes in source depth and volume of magmatism on Mercury
  and their observational signatures}, \textit{Nature Communications},
  \textit{8}, 1945, \doi{10.1038/s41467-017-01692-0}.

\bibitem[{\textit{{Peplowski} et~al.}(2014)\textit{{Peplowski}, {Evans},
  {Stockstill-Cahill}, {Lawrence}, {Goldsten}, {McCoy}, {Nittler}, {Solomon},
  {Sprague}, {Starr}, and {Weider}}}]{peplowski2014}
{Peplowski}, P.~N., L.~G. {Evans}, K.~R. {Stockstill-Cahill}, D.~J. {Lawrence},
  J.~O. {Goldsten}, T.~J. {McCoy}, L.~R. {Nittler}, S.~C. {Solomon}, A.~L.
  {Sprague}, R.~D. {Starr}, and S.~Z. {Weider} (2014), {Enhanced sodium
  abundance in Mercury's north polar region revealed by the MESSENGER Gamma-Ray
  Spectrometer}, \textit{Icarus}, \textit{228}, 86--95,
  \doi{10.1016/j.icarus.2013.09.007}.

\bibitem[{\textit{{Peplowski} et~al.}(2015)\textit{{Peplowski}, {Lawrence},
  {Feldman}, {Goldsten}, {Bazell}, {Evans}, {Head}, {Nittler}, {Solomon}, and
  {Weider}}}]{peplowski2015}
{Peplowski}, P.~N., D.~J. {Lawrence}, W.~C. {Feldman}, J.~O. {Goldsten},
  D.~{Bazell}, L.~G. {Evans}, J.~W. {Head}, L.~R. {Nittler}, S.~C. {Solomon},
  and S.~Z. {Weider} (2015), {Geochemical terranes of Mercury's northern
  hemisphere as revealed by MESSENGER neutron measurements}, \textit{Icarus},
  \textit{253}, 346--363, \doi{10.1016/j.icarus.2015.02.002}.

\bibitem[{\textit{{Phillips} et~al.}(2018)\textit{{Phillips}, {Byrne}, {James},
  {Mazarico}, {Neumann}, and {Perry}}}]{phillips2018}
{Phillips}, R.~J., P.~K. {Byrne}, P.~B. {James}, E.~{Mazarico}, G.~A.
  {Neumann}, and M.~E. {Perry} (2018), Mercury's crust and lithosphere:
  Structure and mechanics, in \textit{Mercury: The View after MESSENGER},
  edited by S.~C. {Solomon}, L.~R. {Nittler}, and B.~J. {Anderson}, pp. 52--84,
  Cambridge University Press, Cambridge, \doi{10.1017/9781316650684.004}.

\bibitem[{\textit{{Rivoldini} and {Van Hoolst}}(2013)}]{rivoldini2013}
{Rivoldini}, A., and T.~{Van Hoolst} (2013), {The interior structure of Mercury
  constrained by the low-degree gravity field and the rotation of Mercury},
  \textit{Earth Planet. Sci. Lett.}, \textit{377}, 62--72,
  \doi{10.1016/j.epsl.2013.07.021}.

\bibitem[{\textit{{Roberts} and {Barnouin}}(2012)}]{roberts2012}
{Roberts}, J.~H., and O.~S. {Barnouin} (2012), {The effect of the Caloris
  impact on the mantle dynamics and volcanism of Mercury}, \textit{J. Geophys.
  Res.}, \textit{117}(E2), E02007, \doi{10.1029/2011JE003876}.

\bibitem[{\textit{{Smith} et~al.}(2012)\textit{{Smith}, {Zuber}, {Phillips},
  {Solomon}, {Hauck}, {Lemoine}, {Mazarico}, {Neumann}, {Peale}, {Margot},
  {Johnson}, {Torrence}, {Perry}, {Rowlands}, {Goossens}, {Head}, and
  {Taylor}}}]{smith2012}
{Smith}, D.~E., M.~T. {Zuber}, R.~J. {Phillips}, S.~C. {Solomon}, S.~A.
  {Hauck}, F.~G. {Lemoine}, E.~{Mazarico}, G.~A. {Neumann}, S.~J. {Peale},
  J.-L. {Margot}, C.~L. {Johnson}, M.~H. {Torrence}, M.~E. {Perry}, D.~D.
  {Rowlands}, S.~{Goossens}, J.~W. {Head}, and A.~H. {Taylor} (2012), {Gravity
  field and internal structure of Mercury from MESSENGER}, \textit{Science},
  \textit{336}, 214, \doi{10.1126/science.1218809}.

\bibitem[{\textit{{Sori}}(2018)}]{sori2018}
{Sori}, M.~M. (2018), {A thin, dense crust for Mercury}, \textit{Earth Planet.
  Sci. Lett.}, \textit{489}, 92--99, \doi{10.1016/j.epsl.2018.02.033}.

\bibitem[{\textit{{Stockstill-Cahill} et~al.}(2012)\textit{{Stockstill-Cahill},
  {McCoy}, {Nittler}, {Weider}, and {Hauck}}}]{stockstillcahill2012}
{Stockstill-Cahill}, K.~R., T.~J. {McCoy}, L.~R. {Nittler}, S.~Z. {Weider}, and
  I.~{Hauck}, Steven~A. (2012), {Magnesium-rich crustal compositions on
  Mercury: Implications for magmatism from petrologic modeling}, \textit{J.
  Geophys. Res.}, \textit{117}, E00L15, \doi{10.1029/2012JE004140}.

\bibitem[{\textit{{Tosi} and {Padovan}}(2020)}]{tosi2019}
{Tosi}, N., and S.~{Padovan} (2020), {Mercury, Moon, Mars: Surface expressions
  of mantle convection and interior evolution of stagnant-lid bodies}, in
  \textit{Mantle Convection and Surface Expressions}, edited by H.~{Marquardt},
  M.~{Ballmer}, S.~{Cottar}, and K.~{Jasper}, p. arXiv:1912.05207, AGU
  Monograph Series, in press.

\bibitem[{\textit{{Tosi} et~al.}(2013)\textit{{Tosi}, {Grott}, {Plesa}, and
  {Breuer}}}]{tosi2013}
{Tosi}, N., M.~{Grott}, A.~C. {Plesa}, and D.~{Breuer} (2013), {Thermochemical
  evolution of Mercury's interior}, \textit{J. Geophys. Res.},
  \textit{118}(12), 2474--2487, \doi{10.1002/jgre.20168}.

\bibitem[{\textit{{Tosi} et~al.}(2015)\textit{{Tosi}, {{\v C}adek}, {B{\v
  e}hounkov{\'a}}, {K{\'a}{n}ov{\'a}}, {Plesa}, {Grott}, {Breuer}, {Padovan},
  and {Wieczorek}}}]{tosi2015}
{Tosi}, N., O.~{{\v C}adek}, M.~{B{\v e}hounkov{\'a}}, M.~{K{\'a}{n}ov{\'a}},
  A.-C. {Plesa}, M.~{Grott}, D.~{Breuer}, S.~{Padovan}, and M.~A. {Wieczorek}
  (2015), {Mercury's low-degree geoid and topography controlled by
  insolation-driven elastic deformation}, \textit{Geophys. Res. Lett.},
  \textit{42}, 7327--7335, \doi{10.1002/2015GL065314}.

\bibitem[{\textit{{Vander Kaaden} et~al.}(2017)\textit{{Vander Kaaden},
  {McCubbin}, {Nittler}, {Peplowski}, {Weider}, {Frank}, and
  {McCoy}}}]{vanderkaaden2017}
{Vander Kaaden}, K.~E., F.~M. {McCubbin}, L.~R. {Nittler}, P.~N. {Peplowski},
  S.~Z. {Weider}, E.~A. {Frank}, and T.~J. {McCoy} (2017), {Geochemistry,
  mineralogy, and petrology of boninitic and komatiitic rocks on the mercurian
  surface: Insights into the mercurian mantle}, \textit{Icarus}, \textit{285},
  155--168, \doi{10.1016/j.icarus.2016.11.041}.

\bibitem[{\textit{Wasylenki et~al.}(2003)\textit{Wasylenki, Baker, Kent, and
  Stolper}}]{wasylenki2003}
Wasylenki, L.~E., M.~B. Baker, A.~J.~R. Kent, and E.~M. Stolper (2003),
  {Near-solidus melting of the shallow upper mantle: Partial melting
  experiments on depleted peridotite}, \textit{J. Petrol.}, \textit{44}(7),
  1163--1191, \doi{10.1093/petrology/44.7.1163}.

\bibitem[{\textit{{Weider} et~al.}(2015)\textit{{Weider}, {Nittler}, {Starr},
  {Crapster-Pregont}, {Peplowski}, {Denevi}, {Head}, {Byrne}, {Hauck}, {Ebel},
  and {Solomon}}}]{weider2015}
{Weider}, S.~Z., L.~R. {Nittler}, R.~D. {Starr}, E.~J. {Crapster-Pregont},
  P.~N. {Peplowski}, B.~W. {Denevi}, J.~W. {Head}, P.~K. {Byrne}, S.~A.
  {Hauck}, D.~S. {Ebel}, and S.~C. {Solomon} (2015), {Evidence for geochemical
  terranes on Mercury: Global mapping of major elements with MESSENGER's X-Ray
  Spectrometer}, \textit{Earth Planet. Sci. Lett.}, \textit{416}, 109--120,
  \doi{10.1016/j.epsl.2015.01.023}.

\bibitem[{\textit{{Wieczorek} and {Meschede}}(2018)}]{wieczorek2018a}
{Wieczorek}, M.~A., and M.~{Meschede} (2018), {SHTools -- Tools for working
  with spherical harmonics}, \textit{Geochem. Geophys. Geosyst.},
  \doi{10.1029/2018GC007529}.

\bibitem[{\textit{{Wieczorek} and {Phillips}}(1998)}]{wieczorek1998}
{Wieczorek}, M.~A., and R.~J. {Phillips} (1998), {Potential anomalies on a
  sphere - Applications to the thickness of the lunar crust}, \textit{J.
  Geophys. Res.}, \textit{103}, 1715--1724, \doi{10.1029/97JE03136}.

\bibitem[{\textit{{Wieczorek} and {Simons}}(2005)}]{wieczorek2005}
{Wieczorek}, M.~A., and F.~J. {Simons} (2005), {Localized spectral analysis on
  the sphere}, \textit{Geophys. J. Int.}, \textit{162}, 655--675,
  \doi{10.1111/j.1365-246X.2005.02687.x}.

\bibitem[{\textit{{Wieczorek} et~al.}(2013)\textit{{Wieczorek}, {Neumann},
  {Nimmo}, {Kiefer}, {Taylor}, {Melosh}, {Phillips}, {Solomon},
  {Andrews-Hanna}, {Asmar}, {Konopliv}, {Lemoine}, {Smith}, {Watkins},
  {Williams}, and {Zuber}}}]{wieczorek2013}
{Wieczorek}, M.~A., G.~A. {Neumann}, F.~{Nimmo}, W.~S. {Kiefer}, G.~J.
  {Taylor}, H.~J. {Melosh}, R.~J. {Phillips}, S.~C. {Solomon}, J.~C.
  {Andrews-Hanna}, S.~W. {Asmar}, A.~S. {Konopliv}, F.~G. {Lemoine}, D.~E.
  {Smith}, M.~M. {Watkins}, J.~G. {Williams}, and M.~T. {Zuber} (2013), {The
  crust of the Moon as seen by GRAIL}, \textit{Science}, \textit{339},
  671--675, \doi{10.1126/science.1231530}.

\bibitem[{\textit{{Wieczorek} et~al.}(2018)\textit{{Wieczorek}, {Meschede},
  {Sales de Andrade}, {Oshchepkov}, {Xu}, and {Walker}}}]{wieczorek2018b}
{Wieczorek}, M.~A., M.~{Meschede}, E.~{Sales de Andrade}, I.~{Oshchepkov},
  B.~{Xu}, and A.~{Walker} (2018), {SHTOOLS: Version 4.2.}, \textit{Zenodo},
  \doi{110.5281/zenodo.592762}.

\bibitem[{\textit{{Wiggins} et~al.}(2019)\textit{{Wiggins}, {Johnson},
  {Bowling}, {Melosh}, and {Silber}}}]{wiggins2019}
{Wiggins}, S.~E., B.~C. {Johnson}, T.~J. {Bowling}, H.~J. {Melosh}, and E.~A.
  {Silber} (2019), {Impact fragmentation and the development of the deep lunar
  megaregolith}, \textit{J. Geophys. Res.}, \textit{124}(4), 941--957,
  \doi{10.1029/2018JE005757}.

\end{thebibliography}
\renewcommand{\baselinestretch}{0.5}

\end{document}